\documentclass[onecolumn, draftcls]{IEEEtran}

\usepackage{amssymb}
\usepackage{amscd}
\usepackage{amsthm}
\usepackage{amsmath}
\usepackage{latexsym}
\usepackage{graphicx}
\usepackage{delarray}
\usepackage{epic}
\usepackage{float}
\usepackage[dvips]{epsfig}
\usepackage{subfig}

\begin{document}


\newtheorem{theorem}{Theorem}[section]
\newtheorem{claim}{Claim}
\newtheorem{definition}[theorem]{Definition}
\newtheorem{lemma}[theorem]{Lemma}
\newtheorem{proposition}[theorem]{Proposition}
\newtheorem{corollary}[theorem]{Corollary}
\newtheorem{example}[theorem]{Example}
\newtheorem{remark}[theorem]{Remark}

\hfuzz5pt 
\hfuzz-50pt

\newcommand{\gt}{\tilde{g}}
\newcommand{\R}{\mathbb{R}}
\newcommand{\Z}{\mathbb{Z}}
\newcommand{\N}{\mathbb{N}}
\newcommand{\Zt}{\mathbb{Z}^2}
\newcommand{\Zd}{\mathbb{Z}^d}
\newcommand{\Ztd}{0\mathbb{Z}^{2d}}
\newcommand{\Rt}{\R^2}
\newcommand{\Rtd}{\R^{2d}}
\newcommand{\Zp}{Z_p}
\newcommand{\al}{\alpha}
\newcommand{\be}{\beta}
\newcommand{\om}{\omega}
\newcommand{\Om}{\Omega}
\newcommand{\ga}{\gamma}
\newcommand{\de}{\delta}
\newcommand{\la}{\lambda}
\newcommand{\La}{\Lambda}
\newcommand{\ala}{\la^\circ}
\newcommand{\aLa}{\La^\circ}
\newcommand{\nat}{\natural}
\newcommand{\G}{\mathcal{G}}
\newcommand{\A}{\mathcal{A}}
\newcommand{\V}{\mathcal{V}}
\newcommand{\M}{\mathcal{M}}
\newcommand{\MV}{\mathcal{MV}}
\newcommand{\MP}{\mathcal{MP}}
\newcommand{\Sp}{\mathcal{S}}
\newcommand{\Pp}{\mathcal{P}}
\newcommand{\W}{\mathcal{W}}
\newcommand{\Hp}{\mathcal{H}}
\newcommand{\ka}{\kappa}
\newcommand{\cast}{\circledast}
\newcommand{\id}{\mbox{Id}}
\newcommand{\by}{\mathbf{Y}}
\newcommand{\bx}{\mathbf{X}}
\newcommand{\bz}{\mathbf{Z}}
\newcommand{\bn}{\mathbf{N}}
\newcommand{\bv}{\mathbf{v}}
\newcommand{\bu}{\mathbf{u}}
\newcommand{\bA}{\mathbf{A}}
\newcommand{\bC}{\mathbf{C}}
\newcommand{\bD}{\mathbf{D}}
\newcommand{\bh}{\mathbf{h}}
\newcommand{\bff}{\mathbf{f}}
\newcommand{\bH}{\mathbf{H}}
\newcommand{\bV}{\mathbf{V}}
\newcommand{\bU}{\mathbf{U}}
\newcommand{\bX}{\mathbf{x}}
\newcommand{\byy}{\mathbf{y}}
\newcommand{\bZ}{\mathbf{z}}
\newcommand{\bt}{\mathbf{t}}
\newcommand{\bs}{\mathbf{\sigma}}
\newcommand{\bI}{\mathbf{I}}

\newcommand{\lo}{{\ell_1}}
\newcommand{\Lo}{{L_1}}
\newcommand{\Lt}{{L_2}}
\newcommand{\SO}{{S_0}}
\newcommand{\SOc}{{S_{0,c}}}
\newcommand{\Lp}{{L^p}}
\newcommand{\Los}{{L^1_s}}
\newcommand{\WCl}{{W(C_0,\ell^1)}}
\newcommand{\lt}{{\ell_2}}

\newcommand{\conv}[2]{{#1}\,\ast\,{#2}}
\newcommand{\twc}[2]{{#1}\,\nat\,{#2}}
\newcommand{\mconv}[2]{{#1}\,\cast\,{#2}}
\newcommand{\set}[2]{\Big\{ #1 \, \Big| \, #2 \Big\}}
\newcommand{\inner}[2]{\langle #1,#2\rangle}
\newcommand{\dotp}[2]{ #1 \, \cdot \, #2}

\newcommand{\Zpd}{\Zp^d}
\newcommand{\I}{\mathcal{I}}
\newcommand{\J}{\mathcal{J}}
\newcommand{\Zq}{Z_q}
\newcommand{\Zqd}{Zq^d}
\newcommand{\Zak}{\mathcal{Z}_{a}}
\newcommand{\C}{\mathbb{C}}
\newcommand{\F}{\mathcal{F}}

\newcommand{\convL}[2]{{#1}\,\ast_L\,{#2}}
\newcommand{\abs}[1]{\lvert#1\rvert}
\newcommand{\absbig}[1]{\big\lvert#1\big\rvert}
\newcommand{\scp}[1]{\langle#1\rangle}
\newcommand{\norm}[1]{\lVert#1\rVert}
\newcommand{\normmix}[1]{\lVert#1\rVert_{\ell^{2,1}}}
\newcommand{\normbig}[1]{\big\lVert#1\big\rVert}
\newcommand{\normBig}[1]{\Big\lVert#1\Big\rVert}


\title{Sub-Nyquist Sampling of Short Pulses}

\author{Ewa Matusiak, Yonina C. Eldar,~\IEEEmembership{Senior~Member,~IEEE}
\thanks{Copyright (c) 2011 IEEE. Personal use of this material is permitted.
However, permission to use this material for any other purposes must be
obtained from the IEEE by sending a request to pubs-permissions@ieee.org}
\thanks{ The first author is with the department of mathematics,
NuHAG, University of Vienna, Austria (phone: +43-1-427750737, fax: +43-1-427750690, email: ewa.matusiak@univie.ac.at) and the second author is with the department of electrical engineering, Technion--Israel Institute of Technology, Haifa, Israel (phone: +972-4-8294682, fax: +972-4-8295757, e-mail: yonina@ee.technion.ac.il)}}

\maketitle


\begin{abstract}
We develop sub-Nyquist sampling systems for analog signals
comprised of several, possibly overlapping, finite duration pulses with unknown
shapes and time positions.
Efficient sampling schemes when either the pulse shape or the
locations of the pulses are known have been previously developed. To
the best of our knowledge, stable and low-rate sampling strategies for
continuous signals that are superpositions of unknown pulses without knowledge
of the pulse locations have not been derived. The goal in this paper is to fill
this gap. We propose a multichannel scheme based on Gabor
frames that exploits the sparsity of signals in time and enables
sampling multipulse signals at sub-Nyquist rates. Moreover, if the signal is
additionally essentially multiband, then the sampling scheme can be adapted to
lower the sampling rate without knowing the band locations. We show that, with
proper preprocessing, the necessary Gabor coefficients, can be
recovered from the samples using standard methods of compressed sensing. In
addition, we provide error estimates on the reconstruction and analyze the
proposed architecture in the presence of noise.
\end{abstract}


\section{Introduction}\label{sec:intro}

One of the common assumptions in sampling theory suggests that in
order to perfectly reconstruct a bandlimited analog signal from its
samples, it must be sampled at the Nyquist rate, that is twice its
highest frequency. In practice, however, all real life signals are
necessarily of finite duration, and consequently cannot be perfectly
bandlimited, due to the uncertainty principle \cite{FS97}. The
Nyquist rate is therefore dictated by the essential bandwidth of the
signal, that is by the desired accuracy of the approximation: the
higher the rate, meaning the more samples are taken, the better the
reconstruction.

In this paper we are interested in sampling a special class of time limited
signals: signals consisting of a stream of short pulses, referred to as
multipulse signals. Since the pulses occupy only a small portion of the signal
support, intuitively less samples, then those dictated by the essential
bandwidth, should suffice to reconstruct the signal. 

There are two standard approaches in the literature to sample such functions.
One is to acquire pointwise samples and approximate
the signal using Shannon's interpolation formula \cite{BS77},
\cite{BS92}. The reconstruction error can be made sufficiently small with just a finite number of samples, when the signal is sampled dense enough.
However, this strategy results in many pointwise samples that are
zero, leading to unnecessary high rates. The second, is to collect Fourier
samples and approximate the signal using a truncated Fourier series.
However, the Fourier transform does not account for local properties of the
signal, hence this method cannot be used to exploit
signal structure and reduce the sampling rate. Both
strategies require the Fourier transform of the
signal to be integrable and do not take the sparsity of the signal in time into
account. Moreover, exact pointwise samples needed for Shannon's method requires
implementing a very high bandwidth sampling filter. Here we show that these
problems can be alleviated using Gabor frames \cite{G01}.

Gabor samples, which are inner products of a function with shifted and
modulated versions of a chosen window, are a good compromise between exact
pointwise samples and Fourier samples. In particular, we show that all
square-integrable time limited signals, without additional conditions on their
Fourier transforms, can be well approximated by truncated Gabor series.
Furthermore,  Gabor samples, taken with respect to a window
that is well localized in time and frequency, provide information about local
behavior of any square integrable function and reflect the sparsity of a
function either in time or frequency. The price to pay is a slightly greater
number of samples necessary for approximation, that comes with using frames,
namely, overcomplete dictionaries. The use of frames is a result of the fact
that Gabor bases are not well localized in both time and frequency \cite{BHW98}.
In all three approaches (pointwise, Fourier, Gabor) the number of samples
necessary to represent an arbitrary time limited signal is dictated by
the essential bandwidth of the signal and the desired approximation accuracy.

Recently, there has been growing interest in efficient sampling of multipulse
signals \cite{VMB02}, \cite{DVB07}, \cite{GTE10}, \cite{TEF10}. This interest
is
motivated by a variety of different applications such
as digital processing of certain radar signals, which are superpositions of
shifted and modulated versions of a single pulse \cite{DVB07}, \cite{BDB10},
\cite{BGE11}. Another example is ultrasound signals,
that can be modeled by superpositions of shifted versions of a given pulse shape
\cite{TEF10}. Multipulse signals are also prevalent in communication channels,
bio-imaging, and digital processing of neuronal signals. Since the
pulses occupy only a small portion of the signal support,
intuitively less samples should suffice to reconstruct the signal.

Prior works mentioned above assumed that the signal is composed of shifts of a
single known pulse. Such signals are completely characterized by a
finite number of parameters and fall under the class of
finite rate of innovation (FRI) signals introduced in \cite{VMB02}.
The sampling schemes proposed in \cite{GTE10} operate at the minimal
sampling rate required for such signals, determined by the rate of
innovation \cite{VMB02}. In this case without noise, perfect recovery is
possible due to the finite dimensionality of the problem.

In this paper we consider sampling of multipulse signals when neither the pulses
nor their locations are known. The pulses can have arbitrary shapes and
positions, and may overlap. The only knowledge we assume is that our signal is
comprised of $N$ pulses, each of maximal width $W$. Despite the complete lack of
knowledge on the signal shape, we show that using Gabor frames and appropriate
processing, such signals can be sampled in an efficient and robust way, using
far fewer samples than that dictated by the Nyquist rate. The number of samples
is proportional to $WN$, that is, the actual time occupancy. More precisely, we
need about $4\mu^{-1} \Om' WN$ samples, where $\Om'$ is related to the essential
bandwidth of the signal and $\mu \in (0,1)$ is the redundancy of the Gabor frame
used for processing. When the signal is additionally sparse in
frequency with only $S$ essential bands of width no more than $\Om_W$, the
sampling rate can be further reduced. For such signals, we need about
$8\mu^{-1}\Om_W' WNS$ samples, where $\Om_W'$ is related to the width $\Om_W$ of the essential bands of the signal. In contrast, Nyquist-rate sampling in both settings requires about $\Om' \be$ samples, where $\be$ is the signal duration. If the signal occupies only a small portion of its time duration, such that $4\mu^{-1}WN \ll \be$, respectively $2\Om_W' S \ll \Om$, then our scheme results in a substantial gain over Nyquist-rate sampling.

The sampling criteria we consider are: a) minimal sampling rate that allows
almost perfect reconstruction, b) no prior knowledge on the locations or shapes
of the pulses, and c) numerical stability in the presence of mismodeling and
noise. To achieve these goals we combine the well established theory
of Gabor frames \cite{G01} with compressed sensing (CS) methods for multiple
measurement systems \cite{T061}, \cite{CRED05}, \cite{ME08}. Our scheme consists
of a multichannel system that modulates the input signal in each channel
with a parametric waveform, based on a chosen Gabor frame, and
integrates the result over a finite time interval. We show that by a
proper selection of the waveform parameters, the Gabor samples can
be recovered, from which the signal is reconstructed. We also consider the
case in which the signal exhibits additional sparsity in frequency, as is
common in radar signals, and show that using our general scheme the sampling
rate can be further reduced. To recover the signal in this case we solve two
CS problems. We then prove that the proposed system is robust to noise and model
errors, in contrast with techniques based on exact pointwise samples.

Our development follows the philosophy of recent
work in analog CS, termed Xampling, which provides a
framework for incorporating and exploiting structure in analog
signals to reduce sampling rates, without the need for discretization
\cite{MEE09}, \cite{MEDS10}. Xampling combines standard analog sampling methods
with CS digital recovery techniques. A pioneer sub-Nyquist system of this
type is the modulated wideband converter (MWC) introduced in \cite{ME09} based
on the earlier work of \cite{MiEl09}. This scheme targets low rate sampling of
multiband signals. Sub-Nyquist sampling is achieved by
applying modulation waveforms to the analog input prior to uniformly sampling at
the low rate.

Another system that falls into the Xampling paradigm is that of \cite{GTE10}
which treats multipulse signals with a known pulse shape. The proposed sampling
scheme is based on modulation waveforms as in the MWC. However, while in the MWC
the modulations are used to reduce the sampling rate relative to the Nyquist
rate, in \cite{GTE10} the modulations serve to simplify the hardware and improve
robustness.

Gabor frames were recently used to sample short discrete pulses in \cite{CEN10}.
The authors analyzed standard CS techniques for redundant
dictionaries, and applied their results to radar-like signals. Finite discrete
multipulse signals were also treated in \cite{HB09} where the authors
modeled the signals as convolutions of a sparse signal with a sparse filter,
both sparse in the standard basis of $\C^N$. The important difference between
\cite{CEN10}, \cite{HB09} and our work is that the former handles discrete time signals. In contrast, our method
directly reduces the sampling rate of continuous time input signals without the
need for discretization.

The paper is organized as follows. In Section~\ref{sec:problem} we
introduce the notation and basic problem definition. Since the main tool in our
analysis is Gabor frames, in Section~\ref{sec:approx} we recall
basic facts and definitions from Gabor theory and show that
truncated Gabor series provide a good approximation for time limited
functions. Based on this observation, in Section~\ref{sec:sampling},
we introduce a sub-Nyquist sampling scheme for multipulse signals.
In Section~\ref{sec:generalizations} we show that
our system can also be used to efficiently sample radar-like
signals, who are sparse in time and
frequency. Section~\ref{sec:related_work} points out connections to recently
developed sampling methods, while Section~\ref{sec:gen_impl} is devoted to
implementation issues. The important part of our design are Gabor windows,
which we review in Section~\ref{sec:windows}. In particular, we summarize
several methods to generate compactly supported Gabor frames. We
demonstrate our theory by several numerical examples in
Section~\ref{sec:simulations}.


\section{Problem Formulation and Main Results}\label{sec:problem}

\subsection{Notation}

We will be working throughout the paper with the Hilbert space of
complex square integrable functions $\Lt(\R)$, with inner product
\begin{equation*}
\inner{f}{g} = \int_{-\infty}^{\infty} f(t)\overline{g(t)}\,dt \quad
\mbox{for all} \quad f,g\in \Lt(\R)
\end{equation*}
where $\overline{g(t)}$ denotes the complex conjugate of $g(t)$. The
norm induced by this inner product is given by $\norm{f}_2^2 = \inner{f}{f}$.
The Fourier transform of $f\in\Lt(\R)$ is
defined as
\begin{equation*}
\widehat{f}(\om) = \int_{-\infty}^{\infty} f(t) e^{-2\pi i \om t}\,dt
\end{equation*}
and is also square integrable with $\norm{\widehat{f}}_2 =
\norm{f}_2$.

A main tool in our derivations are Gabor frames, which we review in
Section~\ref{sec:gabor}. Two important operators that play a central
role in Gabor theory, are the translation and modulation operators
defined for $x,\om \in \R$ as
\begin{equation*}
T_{x} f(t) := f(t-x)\,,\quad M_{\om} f(t) := e^{2\pi i \om t} f(t)\,,
\end{equation*}
respectively. The composition $M_{\om}\, T_x f(t) = e^{2\pi i \om t}
f(t-x)$ is called a time-frequency shift operator and gives rise
to the short-time Fourier transform. For a fixed window
$g\in\Lt(\R)$, the short time Fourier transform of $f\in\Lt(\R)$ with
respect to $g(t)$ is defined as
\begin{equation*}
V_g f(x,\om):= \inner{f}{M_{\om} T_x g}\,.
\end{equation*}

Many derivations, and especially input-output relations for our
sampling systems, will be presented in the compact form of matrix
multiplications. We denote matrices by boldface capital letters, for
example $\bC$, $\bD$, and vectors by boldface lower case letters, such
as $\bX$, $\bZ$.

Our recovery method relies on CS algorithms. An important notion
in this context is that of the restricted isometry property (RIP). A matrix
$\bC$ is
said to have the RIP of order $S$, if there exists $0\leq \delta<1$ such that
\begin{equation*}
(1-\delta) \norm{\bu}_2^2 \leq \norm{\bC \bu}_2^2
\leq (1+\delta) \norm{\bu}_2^2
\end{equation*}
for all $S-$sparse vectors $\bu$ \cite{C08}.

\subsection{Problem Formulation}

\begin{figure}\centering
\includegraphics[width=\columnwidth]{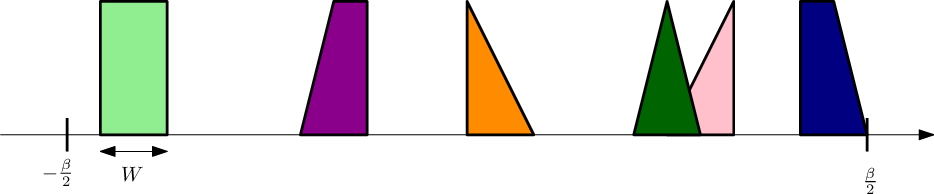}
\caption{Schematic example of a multipulse signal with $N=6$ pulses
each of width no more than $W$. In the example, two of the pulses are
overlapping.}
\label{fig:pulses}
\end{figure}

We consider the problem of sampling and reconstructing signals
comprised of a sum of short, finite duration pulses. A schematic
representation of such a signal is depicted in Fig.~\ref{fig:pulses}.
We do not assume any knowledge of the signal besides the maximum
width (support) of the pulses. More formally, we consider real valued signals
$f(t)$ of the form
\begin{equation}\label{eq:signal model}
f(t) = \sum_{n=1}^N h_n(t)\,, \quad \mbox{where} \quad
\max_{n}{\abs{\mbox{supp}\,h_n}} \leq W\,.
\end{equation}
\noindent The number of pulses $N$ and their maximal width $W$ are
assumed known. The pulses may overlap in time, as in
Fig.~\ref{fig:pulses}. We assume that $f(t)$ is supported on an interval
$[-\be/2,\be/2]$ with $NW \ll \be$. Our goal is to recover $f(t)$ from the
minimal number of samples possible.

Due to the uncertainty principle, finite duration functions cannot
be perfectly bandlimited. However, in practice the main frequency
content is typically confined to a finite interval. We refer to such
signals as essentially bandlimited. More formally, we say that
$f(t)$ is essentially bandlimited, or
$\epsilon_{\Omega}-$bandlimited to $F=[-\Om/2,\Om/2]$, if for some
$\epsilon_{\Omega} < 1$
\begin{equation}\label{def:eMB}
\left ( \int_{F^c}\abs{\widehat{f}(\om)}^2 \, d\om \right )^{1/2} \leq
\epsilon_{\Omega} \norm{f}_2\,.
\end{equation}
The symbol $F^c$ denotes the complement of the set $F$. The
adjective `essential' refers to the fact that the energy of
$\widehat{f}(\om)$ outside $[-\Omega/2,\Omega/2]$ is very small. We
denote the set of multipulse signals (\ref{eq:signal model})
timelimited to $[-\be/2,\be/2]$ and essentially bandlimited to
$[-\Omega/2,\Omega/2]$ by $\MP(N,W,\be,\Om)$.

There are three interesting special cases that fall into the model
(\ref{eq:signal model}). The first is when $h_n(t)$ are shifts of a
known pulse $h(t)$, so that $h_n(t)= \sigma_n h(t-t_n)$ for some
$t_n, \sigma_n \in\R$. In this case, the problem is to find $2N$
parameters, the amplitudes $\sigma_n$ and shifts $t_n$. This setting
can be treated within the class of finite rate of innovation
problems \cite{VMB02}, \cite{GTE10}, \cite{TEF10}.
We return to this scenario in Section~\ref{sec:related_work}
and discuss the relation to our work in more detail. A second class,
is when the location of the pulses $h_n(t)$ are known but the pulses
themselves are not. The third, most difficult scenario, is when
neither the locations nor the pulses are known. Our goal is to
develop an efficient, robust, and low-rate sampling scheme for this
most general scenario. We will later see that our system can be used to
sample signals from the other two cases as well, at their respective
minimal rates.
In Section~\ref{sec:generalizations} we show that our system can be
additionally used to reduce the sampling rate of a special subclass of
$\MP(N,W,\be,\Om)$, which are multipulse signals whose frequency content is
concentrated on only a few bands within $[-\Om/2,\Om/2]$.

We aim at designing a sampling system for signals from the model
$\MP(N,W,\be,\Om)$ that satisfies the following properties:
\begin{itemize}
\item[(i)] the system has no prior knowledge on the locations or shapes of the
pulses;
\item[(ii)] the number of samples should be as low as possible;
\item[(iii)] the reconstruction from the samples should be
simple;
\item[(iv)] the original and reconstructed signals should be close.
\end{itemize}

\subsection{Main Results}

The proposed multichannel sampling method, depicted in
Fig.~\ref{fig:sampling_modulations}, is a mixture of ideas from
Gabor theory and Xampling \cite{MEDS10}. It consists of a set of modulators with functions $q_{j,m}(t)$, followed by integrators over the interval
$[-\be/2,\be/2]$. The system depends on an appropriately chosen Gabor frame with redundancy degree $\mu\in (0,1)$, generated by a compactly supported
window that is well localized in the frequency domain. This frame
provides a sparse representation for $\MP(N,W,\be,\Om)$. The
modulating waveforms $q_{j,m}(t)$, formally defined in
(\ref{eq:q_jm}), are different, finite superpositions of
shifted versions of the chosen Gabor window. The goal of the
modulators is to mix together all windowed pieces of the signal with
different weights, so that, a sufficiently large number of mixtures
will allow to almost perfectly recover relatively sparse multipulse
signals. The resulting samples are weighted superpositions of Gabor coefficients
of the signal with respect to the chosen frame. CS methods
\cite{T061}, \cite{CRED05}, \cite{ME08}
are then used to recover the relevant nonzero signal coefficients
from the given samples.

The number of rows in the resulting CS system is about
$4N\mu^{-1}$; it is a function of the number of pulses present in the signal and
the redundancy $\mu$ of the frame. Since CS algorithms are used
to recover the relevant coefficients, the exact number of rows is dictated by
the RIP constant of the matrix containing the coefficients of the
waveforms, and is given by \textit{O}$(4N\mu^{-1}\log(\be'/(4NW)))$. In the
case of purely multipulse signals, the number of columns is a function of the
desired accuracy of the approximation, and equals about $\Om' W$. However, when
the signal is essentially mutiband, with $S$ bands of width $\Om_W$, then the
number of columns can be reduced to about $2\Om_W' WS$, proportional to the
actual frequency content of the signal. Again, since CS methods are used in
the recovery process, the overall number of columns is dictated by
the RIP constant of the matrix containing the coefficients of the
waveforms, and is given by \textit{O}$(2\Om_W'WS\log(\Om_W'/(2\Om_W'S)))$.
The quantities $\be'$, $\Om'$ and $\Om_W'$ are related to $\be$, $\Om$ and
$\Om_W'$, respectively, and depend on the chosen Gabor frame.

After finding the Gabor coefficients, we recover the signal
using a dual Gabor frame. The function $\widetilde{f}(t)$
reconstructed from the post-processed coefficients satisfies
\begin{equation*}
\norm{f - \widetilde{f}}_2 \leq  \widetilde{C}_0(\epsilon_{\Om} + \epsilon_B )
\norm{f}_2 + \widetilde{C}_1 \norm{\mathbf{n}_1}_2 + \widetilde{C}_2
\norm{\mathbf{n}_2}_2 \,,
\end{equation*}
where $f(t)$ is the original signal, $\widetilde{C}_0$ is a constant
depending on the Gabor frame, and $\epsilon_B$ is related to the
essential bandwidth of the chosen Gabor window. The first term is due to the
signal energy outside the essential bandwidth. The values of
$\mathbf{n}_1$ and $\mathbf{n}_2$ reflect the noise level in the signal
(mismodeling error) and
the samples, respectively, while the constants $\widetilde{C}_1$ and
$\widetilde{C}_2$ depend on the CS method used for
recovery of the Gabor coefficients. If $f(t)$ is perfectly multipulse and the
sampling system is noise free, then $\mathbf{n}_1 = \mathbf{n}_2 =0$. For
multipulse signals that are essentially multiband, $\mathbf{n}_1$ is related to
the signal energy outside the essential bands of $f(t)$.


\section{Sampling using Gabor frames}\label{sec:approx}

We begin by recalling some basic facts and notions from Gabor theory that will
be used throughout the paper, and then show how Gabor frames can be used to
sample multipulse signals with known pulse locations.
In Section~\ref{sec:sampling} we expand the ideas to treat the unknown setting.

\subsection{Basic Gabor Theory}\label{sec:gabor}

A collection $\G(g,a,b) = \{ M_{bl}\,T_{ak} g(t) = e^{2\pi i bl t} g(t-ak)\,;\,
k,l\in\Z \}$ is a Gabor frame
for $\Lt(\R)$ if there exist constants $0<A_1\leq A_2 < \infty$ such
that
\begin{equation*}
A_1 \norm{f}^2 \leq \sum_{k,l\in\Z} \abs{\inner{f}{M_{bl}\,T_{ak}\,g}}^2 \leq
A_2 \norm{f}^2
\end{equation*}
for all $f\in\Lt(\R)$. The frame is called tight, if $A_1 = A_2$. By
simple normalization every tight frame can be changed to a tight
frame with frame bounds equal to one. Therefore, when we talk about
tight frames we will mean frames with frame bounds $A_1=A_2=1$. Every signal
$f\in\Lt(\R)$ can be represented in some Gabor frame \cite{G01}.

A Gabor representation of a signal $f(t)$ comprises the set of
coefficients $\{z_{k,l}\}_{k,l\in\Z}$ obtained by inner products
with the elements of some Gabor system $\G(g,a,b)$ \cite{G01}:
\begin{equation*}
z_{k,l} = \inner{f}{M_{bl}\,T_{ak}\, g} = e^{2\pi ak bl}
\inner{\widehat{f}}{M_{-ak} T_{bl}\,\widehat{g}} \,.
\end{equation*}
The coefficients $z_{k,l}$ are simply samples of a short-time
Fourier transform of $f(t)$ with respect to $g(t)$ at points
$(ak,bl)$. If $\G(g,a,b)$ constitutes a frame for $\Lt(\R)$, then
there exists a function $\ga\in\Lt(\R)$ such that any $f\in\Lt(\R)$
can be reconstructed from $\{z_{k,l} \}_{k,l\in\Z}$
using the formula
\begin{equation}\label{eq:synthesis}
f = \sum_{k,l\in\Z} z_{k,l} M_{bl}\,T_{ak}\,\ga\,.
\end{equation}
The Gabor system $\G(\ga,a,b)$ is the dual frame to $\G(g,a,b)$.
Consequently, the window $\ga(t)$ is referred to as the dual of
$g(t)$. Generally, there is more than one dual window $\ga(t)$. The
canonical dual is given by $\ga=S^{-1}g$, where $S$ is the frame
operator associated with $g(t)$, and is defined by $Sf
=\sum_{k,l\in\Z}\inner{f}{M_{bl}\,T_{ak}\,g} M_{bl}\,T_{ak}\,g$.
There are several ways of finding an inverse of $S$, including the
Janssen representation of $S$, the Zak transform method or
iteratively using one of several available efficient algorithms
\cite{G01}.

Here we will only be working with Gabor frames
whose windows are compactly supported on some interval
$[-\al/2,\al/2]$ and lattice parameters $a=\mu \al$, $b=1/\al$ for
some $\mu\in (0,1)$. For such frames, the frame operator takes on the particularly simple form
\begin{equation*}
S(t) = \sum_{k\in \Z} \abs{g(t-ak)}^2\,.
\end{equation*}
The frame constants can be computed as $A_1 =
\mbox{ess inf } S(t)$ and $A_2 = \mbox{ess sup } S(t)$. The
canonical dual is then $\ga(t) = b S^{-1}(t) g(t)$. For tight frames
the dual atom is simply $\ga(t) = A_1^{-1} b g(t)$. A necessary
condition for $\G(g,a,b)$ to be a frame for $\Lt(\R)$ is that
$ab\leq 1$, while Gabor Riesz bases can only exists if $ab=1$
\cite{G01}. Thus the ratio $1/(ab)$ measures the redundancy of Gabor
systems.

Since one key motivation for considering Gabor frames is to obtain a
joint time-frequency representation of functions one usually
attempts to choose the window $g(t)$ to be well localized in time
and frequency. While the Balian-Low theorem \cite{BHW98} makes it
impossible to design Gabor Riesz bases with good time-frequency
localization, it is not difficult to design Gabor frames with
excellent localization properties. For instance, if $g(t)$ is a
Gaussian, then we obtain a Gabor frame whenever $ab<1$. Therefore,
to obtain a well localized window one needs to allow for certain
redundancy. In Section~\ref{sec:windows} we discuss in detail
how to construct frames and their duals with compactly supported windows based
on \cite{DGM86}, \cite{Ch06}.

We consider windows $g(t)$ that are members of
so-called Feichtinger algebra, denoted by $\SO$ \cite{FZ98}. Such windows
guarantee that the synthesis and analysis mappings
are bounded and consequently result in stable reconstructions, and that the dual
window is in $\SO$. Formally,
\begin{equation*}
\SO := \set{f\in\Lt(\R)}{\norm{V_{\varphi} f}_1 =
\iint \abs{V_{\varphi} f (x,\om)}\,dx\,d\om < \infty}\,,
\end{equation*}
where $\varphi(t) = e^{-\pi t^2}$. The norm in $\SO$ is defined as
$\norm{f}_{\SO}:=\norm{V_{\varphi} f}_1$. Examples of functions
in $\SO$ are the Gaussian, B-splines of positive order, raised
cosine, and any $\Lo(\R)$ function that is bandlimited or any $\Lt(\R)$
function that is compactly supported in time with Fourier transform
in $\Lo(\R)$. Note, that the rectangular window is not a member of $\SO$
since its Fourier transform is not in $\Lo(\R)$.

\subsection{Truncated Gabor Series}\label{sec:truncated_gabor}

It is well known that time limited $\Lt(\R)$ functions, whose Fourier transform
is additionally in $\Lo(\R)$, can be well approximated with a finite number of
samples using a Fourier series. We now show that the same is true for Gabor
series, without assuming anything additional on the signal besides
that it is square integrable.

\begin{figure}\centering
\includegraphics[scale=0.4]{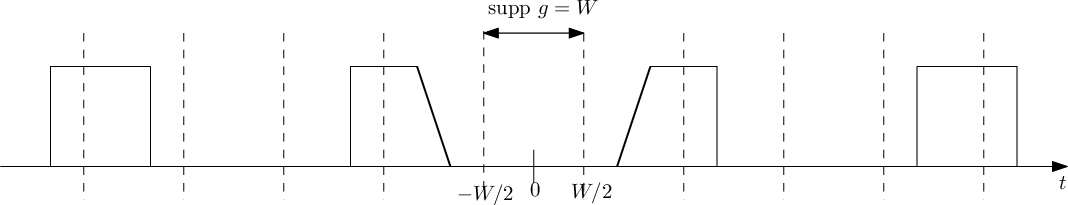}
\caption{The relation between $f$ and the shifts of the support of
$g$ when $\mu=1$. When $\mbox{supp}\,T_{Wk}g$ for some
$k$, does not overlap any of the pulses of $f$, then $z_{k,l}=0$ for
all $\ell$.} \label{fig:shifts}
\end{figure}

Let $\G(g,a,b)$ be a Gabor frame with $g(t)$ compactly supported on
an interval $[-\al/2,\al/2]$, $a=\mu \al$ and $b=1/\al$ for some
$\mu\in (0,1)$. The reason for using compactly supported windows is that for
every function $f(t)$ time limited to $[-\be/2,\be/2]$, the decomposition of
(\ref{eq:synthesis}) reduces to
\begin{equation}\label{eq:gabor sum}
f = \sum_{k=-K_0}^{K_0} \sum_{l\in\Z} z_{k,l} M_{bl}\,T_{ak}\,\ga\,,
\end{equation}
where $\ga(t)$ is a dual window and $K_0$ denotes the smallest
integer such that the sum in (\ref{eq:gabor sum}) contains all
possible non-zero coefficients $z_{k,l}$. The exact value of $K_0$
is calculated by
\begin{equation*}
-\frac{\al}{2} + (K_0+1)a \geq \frac{\be}{2} \Rightarrow
K_0 = \left \lceil \frac{\be + \al}{2a} \right \rceil - 1\,.
\end{equation*}

The number of frequency samples $\ell$ necessary for $\epsilon$ reconstruction
is dictated by the pair of dual windows $(g,\ga)$ as incorporated in the
following theorem, which is an extension of Theorem~3.6.15 in \cite{FZ98}.

\begin{theorem}\label{thm:essband}
Let $f(t)$ be a finite duration signal supported on the interval
$[-\be/2,\be/2]$ and $\epsilon_{\Om}-$bandlimited to
$[-\Om/2,\Om/2]$ and let $\G(g,a,b)$ be a Gabor frame described above with the
dual atom $\ga\in\SO$. Then for every $\epsilon_B >0$ there exists an $L_0 <
\infty$, depending on the dual window $\ga(t)$ and the essential bandwidths
of $g(t)$ and $f(t)$, such that
\begin{equation*}
\normBig{f - \sum_{k=-K_0}^{K_0} \sum_{l=-L_0}^{L_0} z_{k,l}
M_{bl}\,T_{ak}\,\ga}_2 \leq \widetilde{C}_0(\epsilon_{\Om} + \epsilon_B )
\norm{f}_2\,,
\end{equation*}
where $\widetilde{C}_0 = C_{a,b}^2\norm{\ga}_{\SO}\norm{g}_{\SO}$ with
$C_{a,b}=(1+1/a)^{1/2}(1+1/b)^{1/2}$ a constant depending on the
chosen Gabor frame.
\end{theorem}

Similar estimates also appear in \cite{D92}.

\begin{proof}
See Appendix~\ref{proof:thm1}.
\end{proof}

The exact number of frequency coefficients $L_0$ is dictated by the essential
bandwidth of $g(t)$. More precisely, if $g_c(t)$ is a
$[-B/2,B/2]-$bandlimited approximation of $g(t)$ in $\SO$, that is
$\norm{g-g_c}_{\SO}\leq \epsilon_B\norm{g}_{\SO}$, then $L_0 = \left
\lceil \frac{\Om + B}{2b} \right \rceil - 1$.

Theorem~\ref{thm:essband} states that finite duration, essentially
bandlimited signals, can be well approximated using just the dominant
coefficients in the Gabor representation.

The number of samples depend on the chosen frame and the accuracy of the
approximation. To minimize the number of samples for a chosen accuracy of the
approximation, we select $\mu \geq 1/2$ (which reduces the number of
samples in time) and construct a window that is well localized in frequency
(which reduces the number of samples in frequency). Therefore, there is an
interplay between the number of samples in the frequency domain and
the number of samples with respect to time. The total number of Gabor
coefficients, meaning samples of the short-time Fourier transform,
is related to a somewhat larger interval $[-\be'/2,\be'/2] \subseteq
[-\be/2,\be/2]$, where $\be'=\be+\al$, with $K=2K_0+1 \approx \frac{\be'}{a}$,
in the time domain and a larger interval $[-\Om'/2,\Om'/2] \subseteq
[-\Om/2,\Om/2]$, where $\Om'=\Om+B$,
with $L=2L_0+1 \approx \frac{\Om'}{b}$, in the frequency domain. Overall,
the required number of samples is
\begin{align*}
KL &= \left ( 2\left \lceil \frac{\be+\al}{2a} \right \rceil-1 \right ) \left(
2 \left \lceil \frac{\Om+B}{2b} \right \rceil-1 \right )\nonumber \\
&\approx \frac{\be'}{a} \frac{\Om'}{b} = \be'\Om' \mu^{-1}\,.
\end{align*}

When $\mu$ is close to one, and $g(t)$ is well localized in
frequency forming a tight frame, the number of required samples is
close to $\Om\be$. For a fixed $\mu$ and a chosen accuracy of approximation, the
number of frequency samples in a tight frame depends on the decay properties of
$\widehat{g}(\om)$. Therefore, to minimize the number of channels, we need to
choose a window $g(t)$ that exhibits good frequency
localization. On the other hand, having already chosen a frame
$\G(g,a,b)$, if we desire to improve the accuracy of approximation,
then the number $L_0$ of `frequency' coefficients has to increase.

\subsection{Multipulse Signals with Known Pulse
Locations}\label{subsec:MP_known}

If $\al \ll \be$, and the signal is multipulse, then many of the $K$ Gabor
coefficients are zero. Indeed, if the shift $g(t-ak)$
does not overlap any pulse of $f(t)$ then
\begin{equation*}
z_{k,l}= \inner{f}{M_{bl} T_{ak} g} = \int_{-\be/2}^{\be/2} f(t)
\overline{g(t-ak)}e^{-2\pi i blt}\,dt = 0 \,,
\end{equation*}
for all $l\in\Z$. Therefore, when the locations of the pulses
are known, we can reduce the number of samples
from $KL$ to $ML$, where $M<K$ is the number of $k$s, $\abs{k}\leq K_0$, for which $z_{k,l}\neq 0$. To reduce $M$ to minimum, one needs to choose a Gabor frame that allows for the sparsest representation of $f(t)$ with respect to the index $k$.

\begin{figure}\centering
\includegraphics[scale=0.48, trim=0cm -0.2cm 0cm 0cm ]{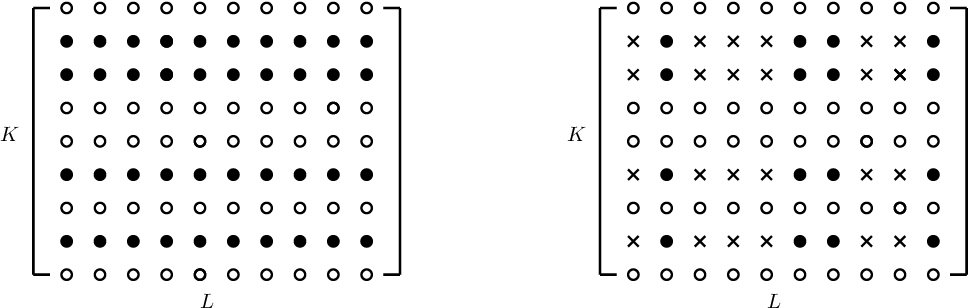}
\caption{Schematic representation of a matrix $\bz$ of Gabor coefficients in the case of multipulse signals, on the left, and multipulse essentially multiband signals, on the right. Empty circles denote zero values and crosses denote small, but nonzero values of $z_{k,l}$.} \label{fig:matrix}
\end{figure}

For signals from $\MP(N,W,\be,\Om)$, an optimal choice is an atom
$g(t)$ that is supported on $[-W/2,W/2]$ and shift parameters $a=\mu
W$, $b=1/W$ for some $\mu \in (0,1)$. In that case at most $\lceil
2\mu^{-1} \rceil$ shifts of $g(t)$ by $ak=\mu W k$ overlap one pulse
of $f(t)$. Indeed, when $\mu=1$, at most two shifts of $g(t)$
overlap one pulse, as depicted in Fig.~\ref{fig:shifts}. When
$\mu<1$, at most $\lceil 2\mu^{-1} \rceil$ shifts of
$\mbox{supp}\,g$ overlap one pulse of $f(t)$. This can be calculated
from
\begin{equation*}
\begin{array}{c} \frac{W}{2} < -\frac{W}{2} +\mu W K_1 \Rightarrow K_1 >
\mu^{-1}\\
-\frac{W}{2} > \frac{W}{2} + \mu W K_2 \Rightarrow K_2 < -\mu^{-1} \end{array}
\Longrightarrow K_1 - K_2 > 2\mu^{-1}\,.
\end{equation*}
Let $\bz$ denote the $K\times L$ matrix of dominant Gabor
coefficients. For functions $f\in
\MP(N,W,\be,\Om)$ each column
$\bz[l]=[z_{-K_0,l},\ldots,z_{K_0,l}]^T$ of $\bz$ has at most
$\lceil 2\mu^{-1} \rceil$ nonzero entries. Moreover, all columns
$\bz[l]$ have nonzero entries at the same places, as modulations
$e^{2\pi i blt}$ applied to $f(t)$ do not change the positions of
the pulses. The matrix $\bz$ is schematically depicted in
Fig.~\ref{fig:matrix}. Therefore, the necessary Gabor coefficients can be
obtained with only $ML \approx 2\Om' W N\mu^{-1}$ channels, where $M=\lceil 2
\mu^{-1}\rceil N$ and $L\approx \Om'W$.

\subsection{Method Comparison}

Since time limited functions can be reconstructed only to a certain
accuracy, we refer to the minimal number of samples as the minimal
number required to reconstruct the signal with a desired
accuracy. For an $\epsilon$ accuracy
of approximation using the Fourier series and Shannon's interpolation
\cite{BS77}, \cite{BS92} methods, the minimal number of samples is of order
$\Om''\be$, where $\Om''$ is such that
\begin{equation*}
\int_{F_1^c} \abs{\widehat{f}(\om)}\,d\om \leq
\epsilon \norm{f}_2\,,
\end{equation*}
with $F_1 = [-\Om''/2,\Om''/2]$ and we have to assume that $\widehat{f}\in
L_1(\R)$. For the above to be satisfied with $\epsilon =
\epsilon_{\Om}$, as in (\ref{def:eMB}), $\Om''$ has to be greater than $\Om$
since $\norm{\widehat{f}}_2 \leq \norm{\widehat{f}}_1$. The approximation error
using Fourier series is then given by
\begin{equation}\label{eq:truncation_error}
\normBig{f(t) - \sum_{\abs{l} \leq L_0} \widehat{f}\left (\frac{l}{\be} \right )
e^{2\pi i lt/\be}}_2 \leq \sum_{\abs{l}> L_0} \abs{\widehat{f}(l/\be)}
\end{equation}
where $\widehat{f}\left (\frac{l}{\be} \right )$ are
the Fourier coefficients and $L_0$ has to be equal at least $\Om'' \be/2$ to
achieve $\epsilon$ approximation. When the signal is multipulse, $L_0$ cannot
be reduced because the Fourier transform does not account for local signal
properties.

The approximation error using Shannon's interpolation formula equals
\begin{equation}\label{eq:shannon}
\left | f(t) - (S_{\Om''}f)(t) \right | \leq
\int_{\abs{\om} > \Om''/2} \abs{\widehat{f}(\om)} \, \,d\om\,,
\end{equation}
where $K_0$ is the largest integer less then $\Om''\be/2$ and
\begin{equation*}
(S_{\Om''}f)(t) = \sum_{\abs{k}\leq K_0} f\left( \frac{k}{\Om''} \right )
\mbox{sinc}(\pi \Om'' (t - k/\Om'') )\,.
\end{equation*}
For $k>K_0$, $f\left( \frac{k}{\Om''} \right )=0$ as $f(t)$ is of finite
duration, so that about $\Om''\be$ pointwise values of $f(t)$ must
be evaluated to achieve $\epsilon$ accuracy. If the signal is multipulse and the pulse locations are known, then this number can be reduced to $NW\Om''$ samples, with $W\Om''$ samples per pulse.

\begin{table}[hbt]\centering
\begin{tabular}{p{1.7cm}||p{1.4cm}|p{1.6cm}|p{2.3cm}}
& \textbf{Fourier \newline series} & \textbf{Shannon's interpolation} &
\textbf{Gabor series with \newline $\G(g,a,b)$, $ab=\mu$} \\ \hline \hline
number of \newline samples & $\approx \Om''\be$  & $\approx \Om''\be$ &
$\approx \Om'\be'\mu^{-1}$ \\ \hline
number of \newline samples for \newline multipulse \newline signal with \newline
known pulse \newline locations & \hspace{1.4cm} \newline \hspace{1.4cm} \newline
$\approx \Om''\be$ & \hspace{1.6cm} \newline \hspace{1.6cm} \newline $\approx
\Om''WN$ & \hspace{2.3cm} \newline \hspace{2.3cm} \newline $\approx 2\Om'WN
\mu^{-1}$ \\
\hline
number of \newline samples for \newline multipulse \newline signal with \newline
unknown pulse \newline locations & \hspace{1.4cm} \newline \hspace{1.4cm}
\newline $\approx \Om''\be$ & \hspace{1.6cm} \newline \hspace{1.6cm} \newline
$\approx \Om''\be$ & \hspace{2.3cm} \newline \hspace{2.3cm} \newline $\approx
4\Om' W N\mu^{-1}$ \\
\hline
approximation \newline error &(\ref{eq:truncation_error}) &
(\ref{eq:shannon}) & Theorem~\ref{thm:essband}
\end{tabular}
\caption{Comparison of three methods for approximating $\Lt(\R)$
functions that are time limited to $[-\be/2,\be/2]$ and essentially
bandlimited to $[-\Om/2,\Om/2]$. The second and third lines refer to
multipulse signals with $N$ pulses, each of width no more than $W$.
The methods are compared for the same accuracy of approximation.}
\label{tab:comparison}
\end{table}

For a Gabor frame with redundancy $\mu$, we achieve $\epsilon$ approximation
with a minimal number of samples of order $\Om'\be'\mu^{-1}$ as long as the
Gabor window $g(t)$ and its dual $\ga(t)$ are such that
\begin{equation*}
\left (\int_{E} \int_{F_2^c} \abs{V_g f(x,\om)}^2\,
dx\,d\om \right )^{1/2}\leq \frac{\epsilon}{C_{a,b} \norm{\ga}_{\SO}}
\norm{f}_2\,,
\end{equation*}
where $E= [-\be'/2,\be'/2]$ and $F_2 = [-\Om'/2,\Om'/2]$. The $\Om'$ is an
enlargement of $\Om$, as in (\ref{def:eMB}), by the essential bandwidth
$[-B/2,B/2]$ of the window $g(t)$, and $\be'$ is an
enlargement of $\be$ by the support $[-W/2,W/2]$ of the window $g(t)$. Then,
$\Om' = \Om + B$ and $\be'=\be+W$.

Table~\ref{tab:comparison} compares the number of samples necessary
for a good approximation of time limited signals and of multipulse signals using
these three methods. As can be seen from the table, the Gabor frame
has two main advantages. The first is that it does not require
strong decay of $\widehat{f}(\om)$ for the reconstruction error to be bounded.
Second, this approach can be used to efficiently sample multipulse signals with
unknown pulse locations, as we will show in the next section. In this case we
need approximately $4\Om' WN \mu^{-1}$ samples which is minimal with
respect to the chosen approximation accuracy and frame redundancy. However, this
amount increases slightly to the order of \textit{O}$(4N\mu^{-1}
\log(\be'/(4NW))\Om'W)$ due to the utilization of CS algorithms in the
recovery process.


\section{Sampling of multipulse signals}\label{sec:sampling}

We now present a sampling scheme for functions from
$\MP(N,W,\be,\Om)$ that reduces the number of channels in
a Gabor sampling scheme and does not require knowledge of the
pulse locations.

\begin{figure*}\centering
\begin{tabular}{cc}
\subfloat[]{\includegraphics[scale=0.65, trim=0cm -0.2cm 0cm
0cm ]{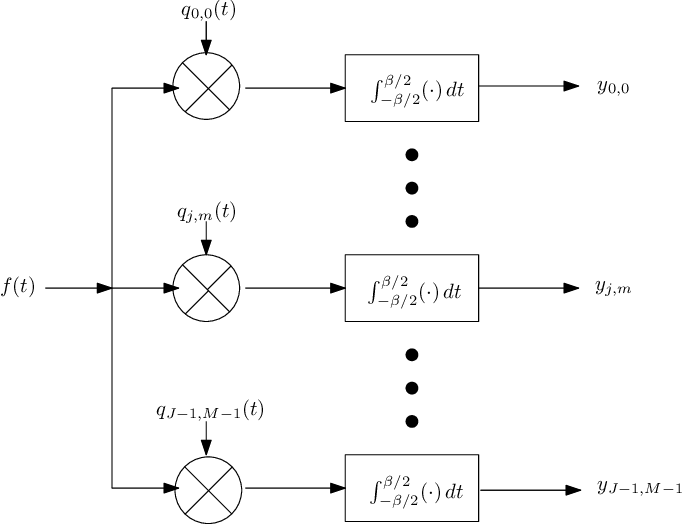}} &
\subfloat[]{\includegraphics[scale=0.65, trim=0cm 0cm
0cm 0cm ]{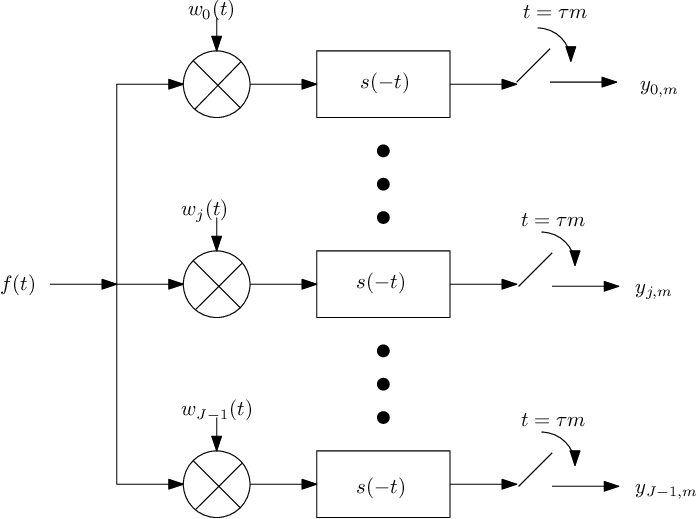}}
\end{tabular}
\caption{An efficient sampling system for multipulse signals (a), and an
equivalent system using filters (b). The sampling step is $\tau=WK$.}
\label{fig:sampling_modulations}
\end{figure*}

\subsection{Sampling System}

Our system, shown in Fig.~\ref{fig:sampling_modulations}(a), exploits
the sparsity of multipulse signals in time. The signal $f(t)$ enters
$JM$ channels simultaneously. In the $(j,m)$th channel, $f(t)$ is
multiplied by a mixing function $q_{j,m}(t)$, followed by an integrator.
The design parameters are the number of channels $JM$ and
the mixing functions $q_{j,m}(t)$, $0\leq m \leq M-1$, $0\leq j \leq
J-1$. The role of the mixing functions is to gather together
all the information in $f(t)$ over the entire interval $[-\be/2,\be/2]$. Namely,
$f(t)$ is windowed with shifts of some compactly supported function, and all
the windowed versions are summed with different weights.

The functions $q_{j,m}(t)$ are constructed from the Gabor frame. Let
$\G(g,a,b)$ be a Gabor frame with window $g(t)$ supported on the
interval $[-W/2,W/2]$, essentially bandlimited to $[-B/2,B/2]$, and
with sampling parameters $a=\mu W$ and $b=1/W$ for some $0< \mu <1$.
Then 
\begin{equation}\label{eq:q_jm}
q_{j,m}(t) = w_j(t) s_m(t)\,,
\end{equation}
where
\begin{eqnarray}\label{eq:w_j/s_m}
w_j(t) &=& \sum_{l=-L_0}^{L_0} d_{jl} e^{-2\pi i bl t} \,, \nonumber \\
s_m(t) &=& \sum_{k=-K_0}^{K_0} c_{mk} \overline{g(t-ak)}\,,
\end{eqnarray}
with
\begin{equation}\label{eq:K_0/L_0}
K_0 = \left \lceil \frac{\be + W}{2W\mu} \right
\rceil - 1 \quad \mbox{and} \quad L_0 = \left \lceil
\frac{(\Om + B)W}{2} \right \rceil - 1\,.
\end{equation}
Let $K=2K_0+1$ and $L=2L_0+1$. The waveforms $q_{j,m}(t)$ are basically mixtures of $KL$ channels $e^{-2\pi i bl t} \overline{g(t-ak)}$, $k=-K_0,\ldots,K_0$ and $l=-L_0,\ldots, L_0$, of the Gabor sampling scheme,
where the functions $w_j(t)$ mix the frequency content of the signal, while $s_m(t)$ mix the temporal content of the signal. To specify
$q_{j,m}(t)$ completely, it remains to choose the coefficients $d_{jl}$ and
$c_{mk}$ defining the waveforms $w_j(t)$ and $s_m(t)$, respectively. To do so,
we first analyze the relation between the samples $y_{j,m}$ and the
signal $f(t)$.

Consider the $(j,m)$th channel:
\begin{eqnarray}\label{eq:samples}
y_{j,m} &=& \int_{-\be/2}^{\be/2} f(t) q_{j,m}(t)\,dt \nonumber \\
&=& \sum_{l=-L_0}^{L_0} d_{jl} \sum_{k=-K_0}^{K_0} c_{mk} \inner{f}{M_{bl}
T_{ak} g} \nonumber \\
&=& \sum_{l=-L_0}^{L_0} d_{jl} \sum_{k=-K_0}^{K_0} c_{mk} z_{k,l}\,.
\end{eqnarray}
The relation (\ref{eq:samples}) ties the known $y_{m,l}$ to the unknown Gabor
coefficients $z_{k,l}$ of $f(t)$ with respect to $\G(g,a,b)$. This relation is
key to the recovery of $f(t)$. If we can recover $z_{k,l}$ from the
samples $y_{j,m}$, then by Theorem~\ref{thm:essband} we are able to
recover $f(t)$ almost perfectly. As can be seen from (\ref{eq:samples}),
the goal of the modulator $q_{j,m}(t)$ is to create mixtures of the
unknown Gabor coefficients $z_{k,l}$. These mixtures, when chosen
appropriately, will allow to recover $z_{k,l}$ from a small
number $JM$ of samples by exploiting their sparsity and
relying on ideas of CS. Note, that when using the basic Gabor
scheme, each $y_{j,m}$ is equal to one value of $z_{k,l}$, so that no combinations are obtained. When $z_{k,l}$ are sparse, with unknown sparsity locations, we will need to acquire all their values using this approach. In contrast, obtaining mixtures of $z_{k,l}$, allows reduction in the number of
samples.

\subsection{Signal Recovery}

For our purposes, it is convenient to write (\ref{eq:samples}) in
matrix form as
\begin{equation}\label{eq:matrix}
\by = \bD \bx^T\,,\quad \mbox{with} \quad \bx=\bC \bz\,.
\end{equation}
Let the indices $k=-K_0,\ldots,K_0$, $l=-L_0,\ldots,L_0$, $m=0,\ldots,M-1$ and
$j=0,\ldots,J-1$ be fixed throughout the exposition. Then, $\by$ is a matrix of
size $J\times M$ whose $jm$th element equals $y_{j,m}$, and $\bx$ is a matrix of
size $M\times L$ with $ml$th element equal $x_{m,l}$. The unknown Gabor
coefficients are gathered in the $K\times L$ matrix $\bz$ with columns $\bz[l]=
[z_{-K_0,l},\ldots,z_{K_0,l}]^T$. The $M\times K$ matrix $\bC$ contains the
coefficients $\bC_{m,k+K_0}=c_{mk}$, while the $J\times L$ matrix $\bD$ contains the coefficients $\bD_{j,l+L_0}=d_{jl}$. The matrices $\bC$ and $\bD$ have to be chosen such that it is possible to retrieve $\bz$ from (\ref{eq:matrix}). If $J=L$, $M=K$ and $\bD$, $\bC$ are identity matrices, then the system of Fig.~\ref{fig:sampling_modulations}(a) reduces to standard Gabor sampling.

From (\ref{eq:samples}) it follows that the waveforms $s_m(t)$,
respectively matrix $\bC$, mix the temporal content of the signal, while
the waveforms $w_j(t)$, respectively matrix $\bD$, mix the frequency content of
the signal. The matrix $\bC$ is used to reduce the number of channels. On the other hand, the purpose of $\bD$ depends on which kind of signals are sampled. For general multipulse signals, the matrix $\bD$ is only used to simplify hardware implementation, as we discuss below, but not to reduce sampling rate. Therefore, in general, we can choose $\bD=\bI$ in this case. For multipulse signals that are additionally frequency sparse, we need $\bD$ to allow recovery
from lower rate samples, namely we can reduce the sampling rate by
using appropriate mixtures with $J<L$, as shown in
Section~\ref{sec:generalizations}.

We begin the discussion with general multipulse signals. When there is no
frequency sparsity, we can choose $J=L$ and $\bD=\bI$, reducing (\ref{eq:samples}) to $\bx = \bC \bz$. In this case $w_j(t)$ become pure
modulations $e^{-2\pi t b(j-L_0)}$. Choosing $J\geq L$ and $\bD$ left invertible leads to a mixture of pure modulations, which can be easier to implement in hardware. This point is discussed in more detail in Section~\ref{sec:gen_impl}. Assuming $\bD$ has full
column rank, we can recover $\bx$ from the samples $\by$ by
$\bx = (\bD^{\dagger} \by)^T$, where $\bD^{\dagger} = (\bD^H \bD)^{-1}\bD^H$ is
the (Moore-Penrose) pseudoinverse of $\bD$. It remains to
retrieve the unknown Gabor coefficients $z_{k,l}$ from $\bx = \bC \bz$.

Recall from Fig.~\ref{fig:matrix}, that for every $\ell$, the
column vectors $\bz[l]$ of matrix $\bz$ have only $\lceil 2\mu^{-1} \rceil N$
out of $K$ nonzero entries, where the nonzero entries correspond to the pulse
locations. In addition, all $\bz[l]$ have nonzero entries in the same
rows. The problem of recovering such a matrix $\bz$ is referred to in the CS
literature as a multiple measurement vector (MMV) problem. Several algorithms have been developed that exploit this structure to recover $\bz$ efficiently from $\bx$ in polynomial time when $\bC$ has the RIP property of order $2\lceil 2\mu^{-1} \rceil N$, twice the number of
nonzero rows \cite{T061}, \cite{CRED05}, \cite{ME08},
\cite{ChHu06}, \cite{DE10}, \cite{EM09}. For example, a popular approach is
by solving  the convex problem
\begin{equation}\label{def:min_1}
\min_{\bz} \norm{\bz}_{2,1} \quad \mbox{subject to} \quad \bx = \bC \bz\,,
\end{equation}
where $\norm{\bz}_{2,1} = \sum_{k=-K_0}^{K_0} (\sum_{l=-L_0}^{L_0}
\abs{z_{k,l}}^2)^{1/2}$.

It is well known that Gaussian and Bernoulli
random matrices, whose entries are drawn independently with equal
probability, have the RIP of order $S$ if $M \geq c S \log (K/S)$,
where $c$ is a constant \cite{BDDW07}, \cite{MPT08}. For random
partial Fourier matrices the respective condition is $M \geq c S
\log^4(K)$ \cite{CT06}, \cite{RV08}. Therefore in our case, the number of
samples in time has to be at least $M \geq 2\lceil 2\mu^{-1} \rceil N \log
(K/(2\lceil 2\mu^{-1} \rceil N))$.

\subsection{Equivalent Representation}

\begin{figure}\centering
\includegraphics[scale=0.5, trim=0cm 0cm -1cm
0cm]{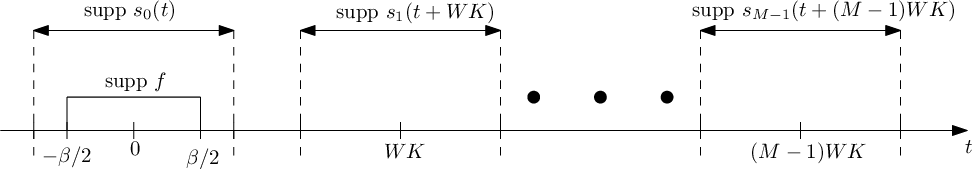} \caption{Relation between the support of the
filter $s(t)$, which is the sum of the shifted supports of $s_m(t)$,
and the support of the signal $f(t)$.} \label{fig:filter-signal}
\end{figure}

For a fixed Gabor frame $\G(g,a,b)$, the number of branches can be reduced to
$J$ if instead of $JM$ modulations followed by an integrator, we perform
$J$ modulations followed by a filter $s(t)$. Consider the system in
Fig.~\ref{fig:sampling_modulations}(b) with $w_j(t)$ as in
(\ref{eq:w_j/s_m}), $\tau = WK$, where $K=2K_0+1$, and $K_0$ is as in
(\ref{eq:K_0/L_0}), and the filter $s(t)$ given by
\begin{equation*}
s(t) = \sum_{m=0}^{M-1} s_m(t+WK m)\,.
\end{equation*}
Note, that for all $m$, $s_m(t)$ is compactly supported in time on
$[-W/2-\mu W K_0, W/2 + \mu W K_0]$, and that its support contains
the support $[-\be/2,\be/2]$ of $f(t)$. The shifted versions
$s_m(t+WKm)$ have non-overlapping supports as the width of
$\mbox{supp}\,s_m$ is smaller than the shift step $WK$
\begin{equation*}
W(1+2\mu K_0) < W(1+2K_0)=WK\,.
\end{equation*}
The support relation between the filter $s(t)$ and the multipulse
signal $f(t)$ is depicted in Fig.~\ref{fig:filter-signal}.

Under these assumptions, the output of the $j$th channel is
\begin{align*}
y_{j,m} &= (\conv{w_j(t) f(t)}{s(-t)})[WKm] \nonumber \\
&= \sum_{n=0}^{M-1} \sum_{l=-L_0}^{L_0} d_{jl}
\inner{M_{-bl}f}{T_{WK(m-n)}\overline{s_n}}\,.
\end{align*}
The sum is nonzero only when $m-n=0$, because otherwise the support
of $s_n(t)$ shifted by $WK(m-n)$ does not overlap the support of
$f(t)$, as depicted in Fig.~\ref{fig:filter-signal}. Therefore it is
sufficient to sample only at points $t=WKm$ for $m=0,\ldots,M-1$,
leading to
\begin{align*}
y_{j,m} &= \sum_{l=-L_0}^{L_0} d_{jl} \inner{M_{-bl}f}{\overline{s_m}} =
\sum_{l=-L_0}^{L_0} d_{jl} \inner{f}{M_{bl}\overline{s_m}} \nonumber \\
&= \sum_{l=-L_0}^{L_0} d_{jl} \sum_{k=-K_0}^{K_0} c_{mk} z_{k,l}\,, \nonumber
\end{align*}
where $z_{k,l} = \inner{f}{M_{bl}\,T_{ak}\,g}$ are the Gabor
coefficients. Evidently, if the coefficients $c_{km}$ used to build
the blocks $s_m(t)$ of the filter $s(t)$ are the same as
coefficients used to create the waveforms $q_{j,m}(t)$, then the two
systems are equivalent.

\subsection{Noisy Measurements}\label{sec:noisy}

Until now we considered signals that were exactly multipulse and
noise free samples. A more realistic situation is when the measurements are
noisy and/or the signal $f(t)$ is not exactly multipulse, having some energy
leaking outside the pulses. We now show that our sampling scheme
is robust to bounded noise in both the signal and the samples.

We say that a signal $f(t)$ essentially bandlimited to
$[-\Om/2,\Om/2]$, is essentially multipulse with $N$ pulses each of
width no more than $W$, if for some $\delta_W<1$ there exists an
$f_p\in\MP(N,W,\be,\Om)$ such that
\begin{equation*}
\norm{f-f_p}_2 \leq \delta_W \norm{f}_2\,.
\end{equation*}
We assume that the signals are time limited to the interval $[-\be/2,\be/2]$,
meaning that the energy leaks only between the pulses, and denote this class of
signals by $\MP_{ess}(N,W,\be,\Om)$.

Since the energy of $f\in \MP_{ess}(N,W,\be,\Om)$ leaks beyond the
support of the pulses, the column vectors $\bz[l]$, of the $K\times
L$ matrix $\bz$ of dominant coefficients, defined in (\ref{eq:matrix}), are no
longer sparse. Nonetheless, $\bz$ can be well approximated by a sparse
matrix $\bz^{\Sp}$, which consists of $S=\lceil 2\mu^{-1}\rceil N$ rows
of $\bz$ with largest $\lt$ norm, and zeros otherwise, and is referred to as the
best $S-$term approximation of $\bz$. The existence of $\bz^{\Sp}$ is shown in
Appendix~\ref{proof:lemma}.

Assuming now that the sampling system of Fig.~\ref{fig:sampling_modulations}(a)
also has imperfections in the form of noise added to the samples, the
input-output relation can be written as
\begin{equation}\label{eq:matrix_general}
\by = \bD\bx^T + \widetilde{\bn}
\end{equation}
where $\bx=\bC \bz$ with $\bz$ a $K\times L$ matrix of Gabor coefficients and
$\widetilde{\bn}$ is an $J\times M$ noise matrix. With $\bD$ having full
column rank, the relation (\ref{eq:matrix_general}) reduces to $\bx = \bC \bz +
\bn$, where $\bn=\bD^{\dagger} \widetilde{\bn}$. A good $S-$term approximation
of $\bz$ can be obtained by utilizing CS algorithms. Specifically, if $\bC$ has
RIP constant $\delta_{2S} \leq \sqrt{2}-1$ and $\bn$ is bounded, then
\begin{equation}\label{eq:min}
\min{\norm{\bz}_{2,1}}\quad \mbox{ subject to } \quad
\norm{\bC\bz - \bx}_2\leq \norm{\bn}_2
\end{equation}
has a unique $S-$sparse solution $\widetilde{\bz}$ that obeys \cite{EM09}
\begin{equation*}
\norm{\bz - \widetilde{\bz}}_2 \leq C_1 \norm{\bz - \bz^{\Sp}}_{2,1} +
C_2 \norm{\bn}_2\,,
\end{equation*}
where $C_1$ and $C_2$ are constants depending on $\delta_{2S}$.

Finally, following a proof similar to that of Theorem~\ref{thm:essband} it can
be shown that a function synthesized from $\widetilde{\bz}$ is a good
approximation of the original signal $f(t)$:
\begin{align*}
\normBig{f &- \sum_{k=-K_0}^{K_0} \sum_{l=-L_0}^{L_0}
\widetilde{z}_{k,l}M_{bl} T_{ak}\ga}_2 \leq \nonumber \\
&\leq \widetilde{C}_0(\epsilon_B + \epsilon_{\Om})\norm{f}_2 + \widetilde{C}_1
\norm{\bz - \bz^{\Sp}}_{2,1} + \widetilde{C}_2 \norm{\bn}_2
\end{align*}
where $\widetilde{C}_0 = C_{a,b}^2\norm{\ga}_{\SO}\norm{g}_{\SO}$,
$\widetilde{C}_1 = C_{a,b} \norm{\ga}_{\SO}C_1$, $\widetilde{C}_2
= C_{a,b}\norm{\ga}_{\SO} C_2$ and $\bn=\bD^{\dagger} \widetilde{\bn}$.

In particular, if $\bz$ is row sparse, as is the case for $f\in
\MP(N,W,\be,\Om)$, then $\bz=\bz^{\Sp}$ and the error of the
approximation depends only on the noise added to the samples. When
the signal is essentially multipulse, then the error bound depends
on the decay of the coefficients. If that quantity is small, then a good
approximation of $f(t)$ is achieved by synthesizing a signal from the solution
$\widetilde{\bz}$ of (\ref{eq:min}). Note here, that the if the dual window
$\ga(t)$ is compactly supported, then a function reconstructed from the
coefficients $\widetilde{\bz}$ is multipulse.


\section{Time-Frequency sparse signals}\label{sec:generalizations}

We now show that we can further reduce the sampling rate when sampling
multipulse essentially multiband signals. We begin by giving a
formal definition of such signals and describe the structure of their Gabor
coefficients.

\subsection{Multipulse Essentially Multiband Signals}

We say that a signal $f\in\MP(N,W,\be,\Om)$ is essentially multiband with $S$
bands of width no more than $\Om_W$, if for some $\epsilon_W<1$ there exists a
multiband function $f_b$ with $S$ bands, all of width no more then $\Om_W$ such
that
\begin{equation*}
\norm{\widehat{f} - \widehat{f_b}}_2 \leq \epsilon_W \norm{\widehat{f}}_2\,.
\end{equation*}
We denote the set of such signals by $\MP(N,W,\be,S,\Om_W,\Om)$. An example
are radar signals that are superpositions of a finite number of time-shifts and
modulations of one pulse. If the generating pulse is well localized in
frequency, then the signal is approximately sparse in the Gabor transform
domain with respect to a window that decays fast in time and frequency.

Let the Gabor frame be as in Section~\ref{sec:sampling}. If the signal is known
to be essentially multiband, then the nonzero row vectors $\bz[k] =
[z_{k,-L_0},\ldots, z_{k,L_0}]$ have only $\lceil (\Om_W + B)W\rceil S$ out of
$L$ dominant entries, and the dominant entries correspond to the locations of
the essential bands of $f(t)$. Indeed, let $[B_1,B_2]$ be any frequency band of
$f(t)$. Then $B_2-B_1\leq \Om_W$ and there are at most $\lceil (\Om_W+B)W\rceil$
shifts of essential bandwidth $[-B/2,B/2]$ of $\widehat{g}(\om)$ by $bl$
that overlap $[B_1,B_2]$. This can be calculated from
\begin{equation*}
\begin{array}{c} B_2 < -\frac{B}{2} +\frac{L_1}{W} \\
B_1 > \frac{B}{2} + \frac{L_2}{W} \end{array}
\Longrightarrow L_1 - L_2 > (\Om_W + B)W\,.
\end{equation*}
The coefficients for which the shift by $bl$ of $g(t)$ does not overlap any
frequency band of $f(t)$ are nonzero
but small. Since there are altogether $S$ bands present, each $\bz[k]$ is
$\lceil (\Om_W + B)W\rceil S-$dominant and all $\bz[k]$ have dominant entries on
the same columns due to the structure of $f(t)$. This
implies that $\bz$ has at most $\lceil (\Om_W+B)W\rceil S$ dominant
columns, as shown in Fig.~\ref{fig:matrix}.

The $K\times L$ matrix $\bz_\Pp$ with $P=\lceil (\Om_W + B)W\rceil S$
nonzero columns corresponding to the $P$ dominant columns of $\bz$ is referred
to as the best $P-$column approximation of $\bz$. Consequently, a result similar
to that of Lemma~\ref{lem:z-zS} holds with time and frequency interchanged. The
sparsity in time combined with the sparsity of $f(t)$ in frequency allows to
further reduce the number of samples necessary for a good reconstruction.

\subsection{Signal Recovery}

To recover $\bz$ from the measurements in
Fig.~\ref{fig:sampling_modulations}(a), the matrices $\bC$ and $\bD$ have to be
chosen appropriately. If $J\geq L$ and $\bD$ is left invertible, then we are
back to the situation of Section~\ref{sec:sampling}. However, since $\bz$ is
additionally almost sparse with respect to columns, we would like to
reduce $J$.

It is convenient to write the relation (\ref{eq:matrix}) as
\begin{equation*}
\by^T = \bC \bU\,, \quad \mbox{where} \quad \bU = \bz \bD^T\,.
\end{equation*}
The matrix $\bU$ inherits sparsity with respect to rows from the matrix $\bz$,
and therefore has only $\lceil 2\mu^{-1} \rceil N$ out of $K$ nonzero rows,
which are precisely the nonzero rows of $\bz$. When the matrix $\bC$ has the RIP
property of order $\lceil 2\mu^{-1} \rceil N$, then $\bU$ can be efficiently
recovered, for example by solving (\ref{def:min_1}) for a unique solution $\bU$
subject to $\by^T = \bC \bU$.

Next, we use $\bU$ to find a unique $\lceil (\Om_W+B)W \rceil
S-$sparse approximation of $\bz$. Let $\Sp$ be the set of indices of
nonzero rows of $\bU$ and $\bU^{\Sp}$ the matrix built from those rows
of $\bU$ indexed by $\Sp$. If the matrix $\bD$ has RIP constant $\delta_{2P}
\leq \sqrt{2}-1$ with $P=\lceil (\Om_W+B)W \rceil S$, then there exists a unique
$P-$sparse solution $\widetilde{\bV}$ of
\begin{equation*}
\min_{\bV} \norm{\bV}_{2,1} \quad \mbox{subject to} \quad (\bU^{\Sp})^T = \bD
\bV\,.
\end{equation*}
Let $\widetilde{\bz}$ be a $K\times L$ matrix whose $\lceil 2\mu^{-1}
\rceil N$ rows indexed by $\Sp$ equal to $\widetilde{\bV}^T$, and the remaining
entries equal to zero. It then follows that $\widetilde{\bz}$ is proportional to the best $P-$column approximation of $\bz$ in the following sense \cite{EM09}
\begin{equation*}
\norm{\bz - \widetilde{\bz}}_2 \leq C_1 \norm{(\bz - \bz_\Pp)^T}_{2,1}\,,
\end{equation*}
where $C_1$ is a constant depending on $\delta_{2P}$. The requirement on $\bD$
translates to $J\geq 2\lceil (\Om_W +B)W \rceil S$. As opposed to purely
multipulse signals, where it suffices to take $\bD=\bI$, this
choice is not possible here, since it does not satisfy RIP.

The resulting matrix $\widetilde{\bz}$ is a $\lceil 2 \mu^{-1}\rceil N$ row
sparse and $\lceil (\Om_W + B)W \rceil S$ column sparse approximation of $\bz$.
It is important to note, that the solution to the first MMV problem, $\by^T =
\bC \bU$, recovers $\bU$ exactly, since $\bU$ is row sparse, while the solution
to the second MMV problem, $\bU^T = \bD \bz^T$, returns a column sparse matrix
$\widetilde{\bz}$ that is an approximation of $\bz$, which itself is not
strictly column sparse.

Finally, the function reconstructed from the coefficients $\widetilde{z}_{k,l}$ is a good
approximation of the input signal $f(t)$:
\begin{align*}
\normBig{f &- \sum_{k=-K_0}^{K_0} \sum_{l=-L_0}^{L_0}
\widetilde{z}_{k,l}M_{bl} T_{ak}\ga}_2 \leq \nonumber \\
&\leq \widetilde{C}_0(\epsilon_B + \epsilon_{\Om})\norm{f}_2 + \widetilde{C}_1
\norm{(\bz - \bz_\Pp)^T}_{2,1}
\end{align*}
where $\widetilde{C}_0 = C_{a,b}^2 \norm{\ga}_{\SO}\norm{g}_{\SO}$ and
$\widetilde{C}_1 = C_{a,b} \norm{\ga}_{\SO}C_1$. The proof is analogous to the
proof of the error estimate in Theorem~\ref{thm:essband} with appropriate
adjustments.

In the case of known positions of the pulses and bands, the minimal sampling
rate for the desired accuracy of the approximation and a given frame is when $M
= \lceil 2\mu^{-1} \rceil N$ and $J = \lceil (\Om_W +B)W \rceil
S$. In the blind setting, when the locations of the pulses and the bands are
unknown, the sampling rate increases by a factor of four (a factor of two in
each domain), with $M \geq  2\lceil 2\mu^{-1} \rceil N$ and $J \geq 2\lceil
(\Om_W +B)W \rceil S$ required for obtaining a unique solution. Therefore, for
signals from the set $\MP(N,W,\be,S,\Om_W,\Om)$, the number of samples $M$ with
respect to time is the same as for signals from $\MP(N,W,\be,\Om)$, while $J$,
the number of samples with respect to frequency, is reduced from $J\geq L$ to
$J<L$. The overall number of samples is $MJ \approx 8\Om_W' W N S\mu^{-1}$,
where $\Om_W' =\Om_W+B$.


\section{Related work}\label{sec:related_work}

Recently, the ideas of CS have been extended to allow for
sub-Nyquist sampling of analog signals \cite{VMB02}, \cite{GTE10}, \cite{TEF10}, \cite{ME09}, \cite{MiEl09}, \cite{EM09}, \cite{El09}, \cite{GE10}. These works follow the Xampling paradigm, which provides a framework for incorporating and exploiting structure in analog signals without the need for discritization \cite{MEE09}, \cite{MEDS10}. Two of these sub-Nyquist solutions are closely related to our scheme: the first is a sub-Nyquist sampling architecture for multiband signals introduced in \cite{ME09}, while the second is a sampling system for multipulse signals with known pulse shape introduced in \cite{GTE10}. We show, that by choosing different waveforms $q_{j,m}(t)$, the systems of \cite{ME09} and \cite{GTE10} are special cases of the system in Fig.~\ref{fig:sampling_modulations}.

\subsection{The Modulated Wideband Converter}

The concept of using modulation waveforms is
based on ideas presented in \cite{ME09} for a multiband model, which is Fourier
dual to ours: the signals in \cite{ME09} are assumed to be sparse in frequency,
while multipulse signals are sparse in time. More specifically, \cite{ME09}
considers multiband signals whose Fourier transform is concentrated on $N$
frequency bands, and the width of each band is no greater than $B$. The
locations of the bands are unknown in advance. A low rate sampling scheme,
called the modulated wideband converter (MWC), allowing recovery of such signals
at the rate of $4NB$ was proposed in \cite{ME09}; a hardware prototype appears
in \cite{MEDS10}. This scheme consists of parallel channels where in each
channel the input is modulated with a periodic waveform followed by a low-pass
filter and low-rate uniform sampling. The main idea is that in each channel the
spectrum of the signal is scrambled, such that a portion of the energy of all
bands appears at baseband. Therefore, the input to the sampler
contains a mixture of all the bands. Mixing of the frequency bands
in \cite{ME09} is analogous to mixing the Gabor coefficients in our
scheme.

The MWC is equivalent to the system of Fig.~\ref{fig:sampling_modulations}(b)
where the waveforms $w_j(t)$ are $B-$periodic and the filter $s(-t)$ is an ideal
rectangular low pass filter, whose bandwidth is $[-B/2,B/2]$. The
samples are taken at points $t=m/B$, $m\in\Z$. The output of the MWC system is
then a weighted sum of Gabor coefficients with respect to a frame $\G(g_r,a,b)$
where $g_r(t)$ is a sinc function that is bandlimited to $[-B/2,B/2]$ and
$a=1/B$, $b=B$. Thus the samples can be written as
\begin{equation*}
y_{j,m} = \sum_{l=-L_0}^{L_0} d_{jm} z_{m,l}\,,
\end{equation*}
where $L_0=\lceil (\Om+B)/(2B) \rceil-1$. With this frame, for each $m\in\Z$ the
number of nonzero Gabor coefficients $z_{m,l}$ equals at most $2B$, as at most
two shifts of $[-B/2,B/2]$ by $bl= Bl$ overlap one band of the signal.
Therefore, the number of channels $J$ is proportional to the number $N$ of
frequency bands in the signal, and equals $J \geq 4NB$.

The MWC is an ideal system, in the sense that it uses ideal low pass
filters, which in practice are difficult to build, and that the reconstruction
process uses infinitely many samples. Using the Gabor approach we can
generalize the MWC to other, not necessarily ideal filters. Furthermore, the
reconstruction error can be computed when only a finite number of samples is available by using Theorem~\ref{thm:essband} with time and frequency interchanged.

The MWC can be easily extended to other, more redundant frames, with a
cost of increased number of channels $J$. Let $\G(g,a,b)$ be a collection of
Gabor frames with windows $g(t)$ bandlimited to $[-B/2,B/2]$ and constants
$a=1/B$ and $b=\mu B$, for some $\mu \in (0,1)$. For any frame from $\G(g,a,b)$,
the MWC parameters change to $L_0 = \lceil (\Om+B)/(2B\mu) \rceil -1$, the
waveforms $w_j(t)$ have to be $B\mu-$periodic and the filter $s(t) =
g(t)$. Also, the sparsity of Gabor coefficients in frequency reduces to $\lceil
2\mu^{-1}\rceil N$, as $\lceil 2\mu^{-1}\rceil$ shifts of $[-B/2,B/2]$ by $bl =
B \mu l$ overlap one band of $f(t)$. The MWC system associated to this
frame has to have $J \geq 2\lceil 2\mu^{-1} \rceil NB\mu^{-1}$ channels. This is an increase in the number of channels by a factor of $\mu^{-1}$. However,
this increase can reduce the number of time samples necessary for achieving the same reconstruction error as with $\G(g_r,1/B,B)$.

\subsection{Multipulse Signals with Known Pulse Shape}

Another related signal model is that of multipulse signals with known pulse
shapes \cite{VMB02}, \cite{GTE10}, \cite{TEF10}:
\begin{equation}\label{eq:stream_pulses}
f(t) = \sum_{s=1}^S \sigma_s h(t-t_s)
\end{equation}
where $h(t)$ is known and $f(t)$ is supported on $[-\beta/2,\beta/2]$. This
problem reduces to finding the amplitudes $\sigma_s$ and
time delays $t_s$. As shown in \cite{VMB02} the time-delays can be estimated using nonlinear techniques e.g. the annihilating filter method as long
as the number of measurements $L$ satisfies $L \geq 2S$ and the
time-delays are distinct. Once the time delays are known, the
amplitudes can be found via a least squares approach. The
number of channels is motivated by the number of unknown parameters
$(\sigma_s,t_s)$ which equals $2S$.

The Fourier coefficients can be determined from the samples of $f(t)$ using a
scheme similar to that of Fig.~\ref{fig:sampling_modulations}(a) with $J \geq L$
channels. The $J$ modulating waveforms being $q_{j,m}(t) = w_j(t)$ with
$b=1/\be$, and all $s_m(t)$ set to one. In this case, the input-output relation
becomes
\begin{equation*}
\byy = \bD \bff\,,
\end{equation*}
where $\byy$ is a vector of length $J$, $\bD$ is a matrix of size $J\times L$
and $\bff$ is a vector of Fourier coefficients $\widehat{f}(l/\be)$ of $f(t)$ of length $L$. If $w_j(t)$ are designed so that $\bD$ is left invertible, then $\bff = \bD^{\dagger} \byy$. We note here, that the system of \cite{GTE10} is inefficient for our signal model, since it reduces to the Fourier series method, which does not take sparsity in time into account. However, by choosing an appropriate Gabor frame and waveforms $q_{j,m}(t)$, the same scheme of Fig.~\ref{fig:sampling_modulations}(a) can be used both to sample signals from the set $\MP(N,W,\be,\Om)$ as well as that of the form (\ref{eq:stream_pulses}), as shown in the following proposition.

\begin{proposition}
Let $\G(g,a,b)$ be a Gabor frame such that $\sum_{k\in\Z} g(t-ak)=1$ almost
everywhere, and the waveforms $q_{j,m}(t)$ in the sampling scheme of
Fig.~\ref{fig:sampling_modulations}(a) are such that the matrix $\bD$,
is left invertible and the matrix $\bC$ has RIP constant $\delta \leq \sqrt{2}
- 1$ of order $2\lceil 2\mu^{-1} \rceil N$ with one row of ones. Then this
sampling scheme can be used to sample multipulse signals of the form
(\ref{eq:stream_pulses}) supported on $[-\be/2,\be/2]$. The time-delays and
amplitudes of $f(t)$ can be retrieved from $J$ samples as long as $t_s \in
[-W/2,W/2]$, $L>2S$ and the known pulse $h(t)$ in (\ref{eq:stream_pulses}) satisfies $\widehat{h}(bl)\neq 0$ for $\abs{\ell}\leq L_0$.
\end{proposition}

The proof is straightforward.

Example of Gabor windows $g(t)$ that are well localized in time and frequency and form a partition of unity, e.g. $\sum_{k\in\Z} g(t-ak)=1$, are the raised cosine window, or B-splines of positive orders \cite{Ch06}. An example of a matrix $\bC$ with a row of ones is a partial Fourier matrix which is known for its good CS properties \cite{C08}.

To conclude, we have seen that the same hardware can be used to sample signals with known pulses and those from $\MP(N,W,\be,\Om)$. The difference is in
the number of branches used and the processing stage.


\section{Waveform Design}\label{sec:gen_impl}

Hardware implementation of our scheme reduces to implementing the
waveforms $q_{j,m}(t)$. The mixing functions $q_{jm}(t)$ are a product of
$w_j(t)$ and $s_m(t)$ defined in (\ref{eq:w_j/s_m}). The functions $s_m(t)$
are pulse sequence modulations, where the sequences are generated to form a
valid CS matrix. An example is a matrix whose entries are $\pm 1$ drawn
independently and with equal probability.

One method to create the waveforms $w_j(t)$ is to low-pass filter $1/b-$periodic waveforms. More precisely, let
\begin{equation*}
\widetilde{w}_j(t) = \sum_{l\in\Z} \sum_{i=0}^{I-1} \al_j[i]
p \left (t-\frac{i}{bI} - \frac{l}{b} \right )
\end{equation*}
where $p(t)$ is some pulse shape such that $\widetilde{w}(bl)\neq 0$ for
$\abs{\ell}\leq L_0$, and $\al_j[i]$ is a length$-I$
sequence. Since $\widetilde{w}_j(t)$ is $1/b$ periodic it can be expressed as
\begin{equation*}
\widetilde{w}_j(t) = \sum_{l\in\Z} \widetilde{d}_{jl} e^{-2\pi i bl t}\,,
\end{equation*}
for some coefficients $\widetilde{d}_{jl}$. We then filter $\widetilde{w}_j(t)$
by a filter $u(t)$ with frequency response $\widehat{u}(\om)$, designed so
that
\begin{equation*}
\widehat{u}(\om) = \left \{ \begin{array}{ll} 1 &
\om = bl,\,\, \abs{l}\leq L_0\\ 0 & \om = bl,\,\,
\abs{l}\geq L_0\\
\mbox{arbitrary} & \mbox{elsewhere} \end{array} \right.\,,
\end{equation*}
to form the waveforms $w_j(t) = \sum_{l=-L_0}^{L_0} d_{jl} e^{-2\pi i bl t}$
with coefficients $d_{jl} = \widetilde{d}_{jl} \cdot \widehat{u}(bl)$.
The shaping filter frequency response, $\widehat{u}(\om)$, is designed to
transfer only the coefficients with index $\ell=-L_0,\ldots, L_0$, suppressing
all other coefficients.

For the matrix $\bD$, built from the coefficients $d_{jl}$, to be left
invertible a necessary condition is that $J\geq I \geq L$ and the sequences
$\al_j[i]$ are chosen such that the matrix $\bA$, whose $ji$th element is
$\al_j[i]$, has full column rank \cite{GTE10}. For example, if $J=I=L$, then the
rows of $\bA$ can be created from cyclic shifts of one basic sequence. On the
other hand, for a matrix $\bD$ to be a valid CS matrix, meaning to have RIP
property with high probability, the values $\al_j[i]=\pm 1$ are chosen
independently with equal probability and $I\geq L \geq J$ \cite{ME09}.

One example of a pulse modulation scheme is when $J=I=L$, and
\begin{equation*}
w(t) = \left \{ \begin{array}{ll} 1 & t\in \left [ 0, \frac{1}{bI} \right ] \\
0 & t\notin \left [ 0, \frac{1}{bI} \right ]\end{array} \right .\,.
\end{equation*}
The frequency response of this pulse is given by
\begin{equation*}
\widehat{w}(\om) = \frac{1}{bI} e^{-\frac{\pi \om}{bI}} \cdot
\mbox{sinc} \left ( \frac{\om}{bI}\right )\,,
\end{equation*}
so that $\widetilde{w}(bl)\neq 0$ for $\abs{\ell}\leq L_0$. In addition we
choose $\al_j[i]$ as sequences of $\pm 1$s, created from cyclic shifts of
one basic sequence, in a way that yields an invertible matrix $\bA$.
Such rectangular pulses with alternating signs can be easily
implemented in hardware \cite{MEDS10}.


\section{Gabor windows}\label{sec:windows}

The sampling scheme presented in this paper is based on Gabor
frames. We recall here some methods to construct Gabor frames with well
localized windows for a chosen redundancy $\mu$ based on results from
\cite{DGM86} and \cite{Ch06}.

Let $\mu \geq \frac{1}{2}$. A window $g(t)$ that is supported on $[-W/2,W/2]$
and forms a frame with $a=\mu W$ and $b=1/W$ can be constructed from an
everywhere increasing function $h(t)$ such that $h(t)=0$ for $t\leq
0$, and $h(t)=1$ for $t\geq 1$ by
\begin{equation*}
g(t) = \left \{ \begin{array}{ll}
0\,,& t\leq -\frac{W}{2}\,,\\
\left [ h\left ( \frac{t/W + 1/2}{1-\mu} \right ) \right ]^{1/2}, &
t\in\left [-\frac{W}{2} ,-\frac{W\lambda}{2} \right ],\\
1\,, & \abs{t} \leq \frac{W\lambda}{2},\\
\left [ 1 - h\left( \frac{t/W - \lambda/2}{1-\mu} \right )\right ]^{1/2}\,,
& t\in \left [\frac{W\lambda}{2},\frac{W}{2} \right ],\\
0\,, & t \geq \frac{W}{2}\,,
\end{array} \right .
\end{equation*}
where $\lambda = 2\mu -1$, \cite{DGM86}. The function $g(t)$ is non-negative,
has the desired support and equals $1$ on $[-W\lambda/2, W\lambda/2]$. If $h(t)$
is taken to be $2k$ continuously differentiable, than $g(t)$ is $k$ times
continuously differentiable, which implies that $\widehat{g}(\om)$
decays like \textit{o}$(\abs{\om}^{-k})$. The points
$t = \pm W\lambda/2$, where $g(t)$ becomes constant, have been
chosen so that their distance to the furthest edge of $\mbox{supp}\,
g$ is exactly $\mu W$. The frame bounds of such a constructed frame
equal $A_1 = A_2 = 1$ \cite{DGM86}, since
\begin{equation*}
\sum_{k\in \Z} \abs{g(t + k \mu W)}^2 = 1\,.
\end{equation*}
As an example, let $h(t)=\sin(\pi t/2)$ on $[0,1]$ and $\mu=\frac{1}{2}$. Then
\begin{equation*}
g(t) = \left \{ \begin{array}{ll} 0 & \abs{t} \geq W/2,\\
\cos(\pi t/W) & \abs{t} \leq W/2\,.  \end{array} \right .
\end{equation*}

An alternative construction for $\mu\geq 1/2$ was developed in
\cite{L09}, \cite{CKK10}. The method results in spline type
windows $g(t)$ of any order of smoothness that satisfy the partition
of unity criterion. The constructions are made by counting the number of
constraints (in the Ron-Shen duality condition \cite{RS95}, and on
the points where continuity/differentiability is required) and then
searching for polynomials on $[-1,0]$ and on $[0,1]$ of a matching
degree. One example is $\widetilde{g}(t)$ supported on
$[-2/3,2/3]$ and given by
\begin{equation*}
\widetilde{g}(t) = \left \{ \begin{array}{ll} 2+3t & t\in [-2/3,-1/3]\,,\\
1 & \abs{t}\leq 1/3\,,\\
2-3t & t\in [1/3,2/3]\,, \end{array} \right .
\end{equation*}
that forms a frame with $a=1$ and $b=3/4$. It forms a partition of
unity with a shift parameter $a=1$, $\sum_{k\in\Z}
\widetilde{g}(t-k)=1$. The dual window is also supported on
$[-2/3,2/3]$ and is given by
\begin{equation*}
\widetilde{\ga}(t) = \left \{ \begin{array}{ll} -18t^2 -15t -2
& t\in [-2/3,-1/3]\,,\\
1 & \abs{t}\leq 1/3\,,\\
-18t^2+15t-2 & t\in [1/3,2/3]\,. \end{array} \right .
\end{equation*}
Applying dilation by $(\mu W)^{-1}$, with $\mu=3/4$, both to $\widetilde{g}(t)$
and $\widetilde{\ga}(t)$ we obtain a dual pair of windows $g(t)$ and $\ga(t)$
\begin{equation*}
g(t) = \widetilde{g}(t/(\mu W))\quad \mbox \quad
\ga(t)= \widetilde{\ga}(t/(\mu W))
\end{equation*}
that are supported on $[-W/2,W/2]$, and such that $\G(g,\mu W, 1/W)$
forms a frame with frame bounds $A_1=1/2$ and $A_2=1$. Moreover,
$g(t)$ forms a partition of unity with shift parameter $a=\mu W$,
$\sum_{k\in\Z} g(t-\mu Wk)=1$.

Well known, compactly supported Gabor windows are the B-splines. Let $B_N(t)$ be
a spline of order $N$,
\begin{equation*}
B_1(t) = \chi_{1/2}(t)\,, \quad B_{N+1}(t) = (\conv{B_N}{B_1})(t)\,.
\end{equation*}
Then $B_N(t)$ is supported on $[-N/2,N/2]$ and forms a partition of
unity with shift parameter $a=1$. To generate a Gabor frame from
$B_N(t)$ with a window supported on $[-W/2,W/2]$ and lattice
parameters $a=\mu W$, $b=1/W$, such that the window forms a
partition of unity with shift $\mu W$, we need to choose $\mu=1/N$
\cite{DP99}. Then $g(t) = B_N(tN/W)$ is supported on the desired
interval and decays like $(1+\abs{\om})^{-N-\epsilon}$ in the
frequency domain. Note that $\mu$ decreases as the order $N$ of
smoothness of the B-spline is increased. Thus smoother windows can
be obtained only at the cost of a smaller $\mu$. However, already
for $N=3$ we get good concentration properties of $g(t)$. The dual can be
computed by inverting the Gabor frame operator, or by using the method of
\cite{Ch06}.


\section{Simulations}\label{sec:simulations}

\begin{figure*}\centering
\begin{tabular}{ccc}
\subfloat[]{\includegraphics[scale=0.42, trim=1.3cm 0.2cm -0.2cm
0.2cm ]{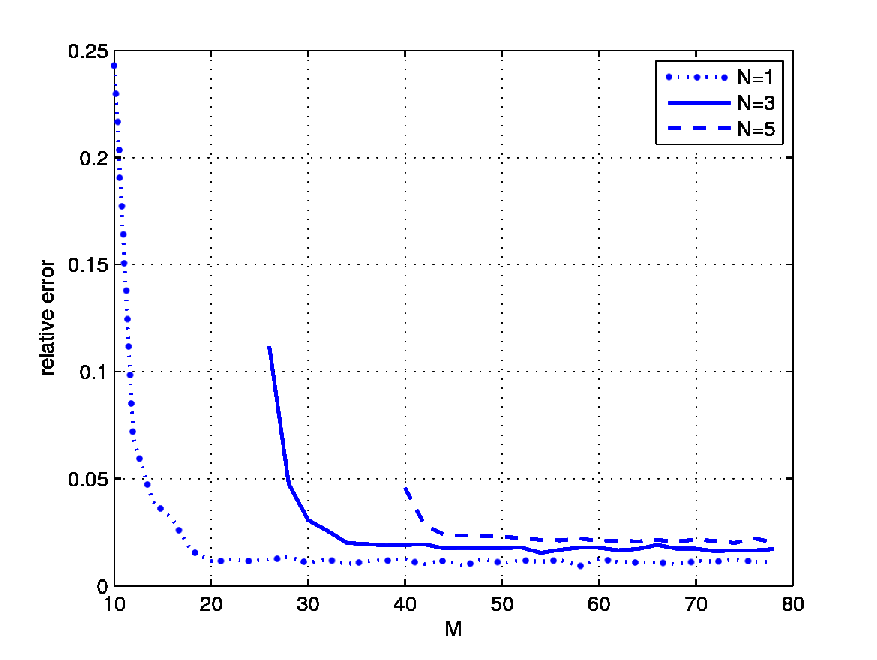}} &
\subfloat[]{\includegraphics[scale=0.42, trim=1.8cm 0.2cm
-0.2cm 0.2cm ]{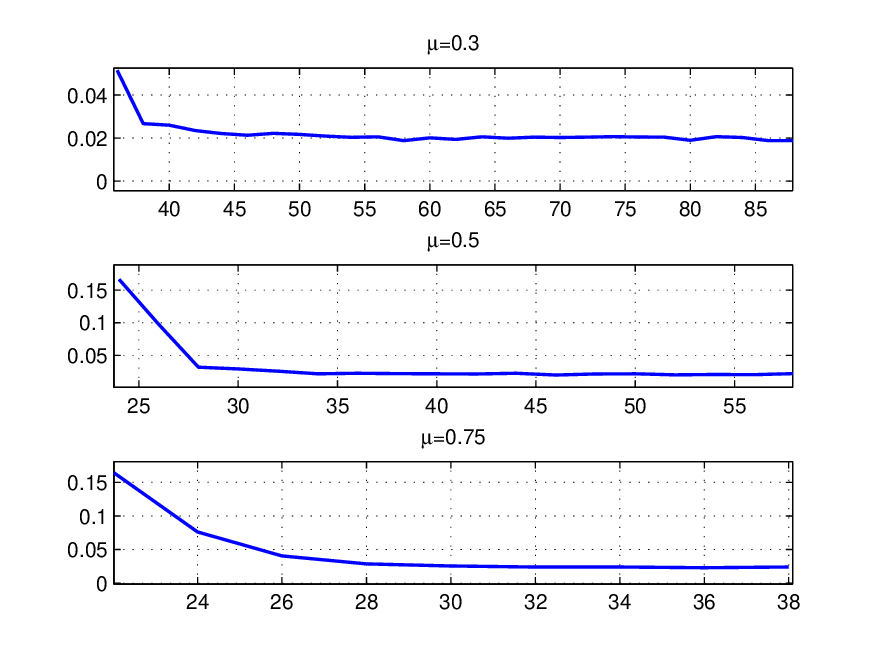}} &
\subfloat[]{\includegraphics[scale=0.42, trim=1.8cm 0.2cm
-0.2cm 0cm ]{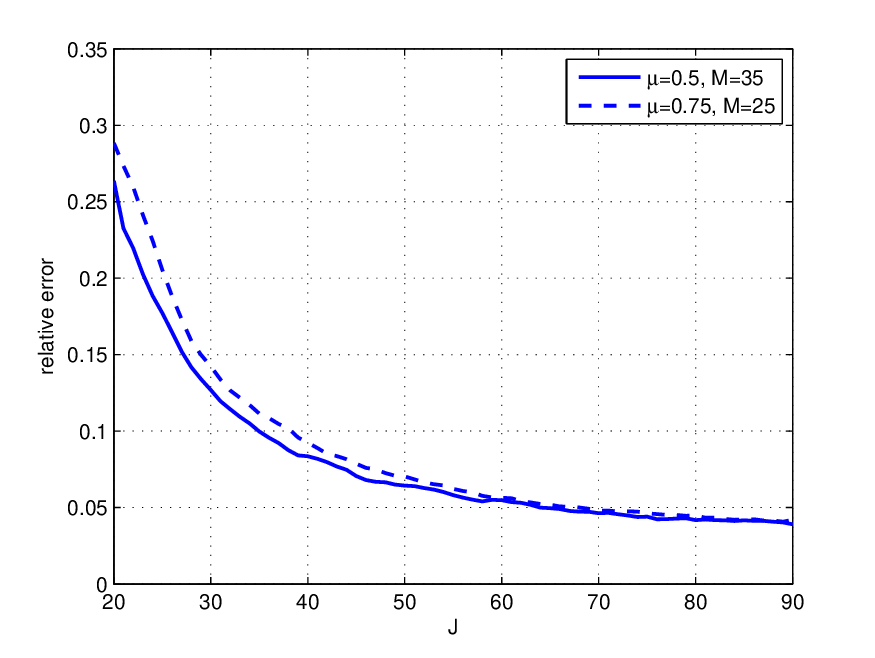}}
\end{tabular}
\caption{(a) Decay of the reconstruction error with increased number of
samples for multipulse signals with $N=1,3,5$ pulses, pulse width $W=0.18$ms and frame redundancy $\mu=0.5$. (b) Comparison of performance for signals with $N=3$, $W=0.24$ms using different frames with $\mu=0.3$, $\mu=0.5$ and $\mu=0.75$. (c) Comparison of the relative error with respect to the number of channels $J$ for multipulse essentially bandlimited signals with $N=3$, $W=0.18$ms, $S=2$ with respect to different frames.}
\label{fig:montecarlo}
\end{figure*}

We now present some numerical experiments illustrating the recovery
of multipulse signals.

We tested our sampling scheme using Monte Carlo simulations averaged over $500$
trials on a range of multipulse signals of duration $\be=22$ms. The pulses making up the signals were chosen at random from a set of five different pulses: cosine, Gaussian, B-spline of order $3$ and $5$, and rectangular pulse. The locations of the pulses were also chosen at random. We varied the number of pulses $N=1,3,5$, the maximal width $W=0.18,0.24$ms of the pulse, and the  redundancy $\mu=0.3,0.5,0.75$ of the frame. Throughout the experiments we chose $\bD=\bI$ and $\bC$ as a Bernoulli random matrix. We measured the relative error $\norm{f-\widetilde{f}}/\norm{f}_2$. For redundancy
$\mu=0.3$ we chose a Gabor frame with window being B-spline of order three, for
$\mu=0.5$ the window was a cosine, and for $\mu=0.75$ we chose the truncated
Gaussian.

Fig.~\ref{fig:montecarlo}(a) depicts the decrease in the reconstruction error with increased number of samples $M$ for different values of $N$ and $W=0.18$ms. We used a tight Gabor frame with a cosine window and redundancy $\mu=0.5$. The $M$ is dictated by the the number of pulses and frame redundancy, and it has to be at least $2\lceil 2\mu^{-1} \rceil N$. Meaning, that for multipulse signals with $N=1$ pulses, $M\geq 8$, for $N=3$ we have $M\geq 24$, and for $N=5$ it has to be $M\geq 40$. As expected, the sparser the signal, the less samples are needed for a good reconstruction. The number of samples in time can be significantly reduced if sparsity is taken into account. Without any knowledge on the sparsity we would have to take $K=241$ time samples for signals with $N=5$ pulses, and with $L_0=5$ that would result in the reconstruction error of $0.05$. However, when sparsity is taken into account, already $M=40$ samples suffice to achieve the same reconstruction error. Therefore reducing the number of samples by a factor of six. When $N=3$, to achieve reconstruction error of $0.05$ we need $M=28$ samples in time, and $M=12$ for signals with $N=1$.

In Fig.~\ref{fig:montecarlo}(b) we considered the influence of the
Gabor frame on the reconstruction error and the number of samples $M$ involved.
We tested the system for signals with $N=3$ pulses of width no more than
$W=0.24$ms and $L_0=5$. The least number of samples $K=121$ is achieved with
$\mu=0.75$ and at the same time with $M=25$ we achieve a good reconstruction.
The value of $M$ necessary for a good reconstruction increases with the increase of redundancy. Without knowing the sparsity structure of the signal in time, we would have to take $K=271$ samples for $\mu=0.3$, and $K=120$ for $\mu=0.5$. When sparsity is exploited, we can reduce that number to $M=45$ and $M=27$, respectively.

We then examined the performance of our sampling scheme on signals comprising
three pulses of width no more than $W=0.18$ms, that are additionally
essentially multiband with two bands. Fig.~\ref{fig:montecarlo}(c)
depicts the decay of reconstruction error with the increase of $J$ for two
different frames: one is a tight frame with cosine window
and redundancy $\mu=0.5$ and second, a frame with Gaussian window and redundancy $\mu=0.75$. The sampling system was tested with the matrix $\bD$ being the random Fourier matrix and $\bC$ a Bernoulli random matrix.
For example, when a frame is of redundancy $\mu=0.5$ and no sparsity is taken
into account then we need $K=241$ and $L=91$ samples to achieve a
reconstruction error of $0.07$. On the other hand, with sparsity being exploited we can use only $M=35$ and $J=40$ for a similar reconstruction quality, resulting in a twelvefold reduction in the number of samples.


\section{Conclusions}

We presented an efficient sampling scheme for multipulse signals,
which is designed independently of the time support of the input
signal. Our system allows to sample multipulse signals at the minimal rate, far
below Nyquist, without any knowledge of the pulse shapes or its locations.
The scheme fits into the broad context of Xampling - a recent
sub-Nyquist sampling paradigm for analog signals. Our architecture relies on
Gabor frames which lead to sparse expansions of multipulse signals, and consists
of modulating the signal with several waveforms followed by integration. We
showed that the Gabor coefficients, necessary for
reconstruction, can be recovered from the samples of the system by
utilizing CS techniques. The number of necessary
samples depends on the desired accuracy of the approximation,
essential bandwidth of the signal, and redundancy factor $\mu$
related to the Gabor frame, and equals $4\Om' NW \mu^{-1}$. The sampling rate
can be further reduced if the signal is additionally sparse in frequency. We
also showed that the proposed sampling and recovery technique is stable with
respect to noise and mismodeling.



\appendices


\section{Proof of Theorem~\ref{thm:essband}}\label{proof:thm1}

The proof is rooted in that of Theorem 3.6.15 in \cite{FZ98} with
appropriate adjustments. Since $\G(g,a,b)$ is a Gabor frame, $f(t)$
admits a decomposition
\begin{equation*}
f = \sum_{k=-K_0}^{K_0} \sum_{l\in\Z} z_{k,l} M_{bl}\,T_{ak}\,\ga\,.
\end{equation*}

Let $\epsilon_B >0$. The bandlimited $\SO$ functions are dense in
$\SO$, therefore, there exists $g_c\in\SO$ bandlimited to some
$[-B/2,B/2]$, such that
\begin{equation*}
\norm{g-g_c}_{\SO} \leq \epsilon_B \norm{g}_{\SO} \,.
\end{equation*}
Since $f(t)$ is an essentially bandlimited function, there exists a
function $f_c(t)$ bandlimited to $[-\Om/2,\Om/2]$, such that
\begin{equation*}
\norm{f-f_c}_2 \leq \epsilon_{\Om} \norm{f}_2\,.
\end{equation*}
Consequently, $\abs{z_{k,l}} =
\abs{\inner{\widehat{f_c}}{M_{-ak}\,T_{bl} \widehat{g_c}}} \neq 0$
only for those $\ell$ such that $\mbox{supp}\, \widehat{f_c} \cap (\mbox{supp}\,
\widehat{g_c} + bl) \neq \emptyset$, that is
\begin{equation*}
[-\Om/2,\Om/2] \cap [bl-B/2,bl+B/2] \neq \emptyset\,.
\end{equation*}
The fact that  $f_c(t)$ and $g_c(t)$ are bandlimited implies that
there are only a finite number of values $\ell$ for which $V_{g_c}
f_c(ak,bl) \neq 0$. Let $L_0$ be the smallest integer such that
$\abs{V_{g_c} f_c(ak,bl)} = 0$ for $\abs{l} > L_0$. The exact value
of $L_0$ can be calculated as
\begin{equation*}
L_0 = \left \lceil \frac{\Om + B}{2b} \right \rceil - 1\,.
\end{equation*}

Define a sequence $d_{k,l}$ as
\begin{equation*}
d_{k,l}=\left \{ \begin{array}{ll} z_{k,l}\,, & \abs{k}\leq K_0,\,\,\abs{l}>
L_0\\
0\,, & \mbox{else.} \end{array} \right .
\end{equation*}
Then $\abs{d_{k,l}} \leq \abs{V_{g-g_c} f (ak,bl) + V_{g_c}
(f-f_c)(ak,bl)}$ for all $k,l\in\Z$, and
\begin{align*}
\normBig{f &- \sum_{k=-K_0}^{K_0} \sum_{l=-L_0}^{L_0} z_{k,l}
M_{bl}\,T_{ak}\,\ga}_2 = \nonumber \\
&= \normBig{\sum_{k\in\Z} \sum_{l\in\Z} d_{k,l}
M_{bl}\,T_{ak}\,\ga}_2 \leq C_{a,b}\norm{\ga}_{\SO} \norm{d}_{\lt} \nonumber \\
&\leq C_{a,b} \norm{\ga}_{\SO} \big (\norm{V_{g-g_c} f}_{\lt} +
\norm{V_{g_c}(f-f_c)}_{\lt} \big ) \nonumber \\
&\leq C_{a,b}^2 \norm{\ga}_{\SO} \norm{g}_{\SO}(\epsilon_B + \epsilon_{\Om})
\norm{f}_2\,
\end{align*}
where we first used the boundedness of the analysis operator related to $g(t)$ and then the synthesis operator related to $\ga(t)$ whenever
$g$ and $\ga$ are in $\SO$.


\section{$S-$term approximation of $\bz$}\label{proof:lemma}

We show here the existence of an $S-$term approximation of $\bz$.

\begin{lemma}\label{lem:z-zS}
Let $f\in \MP_{ess}(N,W,\be,\Om)$ be $\delta_W-$essentially
multipulse and $\G(g,a,b)$ be a Gabor frame with $g$ compactly
supported on $[-W/2,W/2]$ and $a=\mu W$, $b=1/W$ for some $0<\mu <
1$. Then there exists a subset $\Sp$ of $\{-K_0,\ldots,K_0 \}$ such
that
\begin{equation*}
\norm{\bz-\bz^{\Sp}}_{2,1} \leq \delta_W \sqrt{K} C_{a,b}
\norm{g}_{\SO} \norm{f}_2\,,
\end{equation*}
where $\bz^{\Sp}$ consists of rows of $\bz$ indexed by $\Sp$, $K=2K_0+1$ and
$\norm{\bz}_{2,1} = \sum_{k=-K_0}^{K_0}(\sum_{l=-L_0}^{L_0}
\abs{z_{k,l}}^2)^{1/2}$.
\end{lemma}

\begin{proof}
Let $f_p\in\MP(N,W,\be,\Om)$ be a multipulse
$\delta_W-$approximation of $f$. Then $V_g f_p (ak,bl)=0$ for all
$\abs{k}>K_0$, and the column vectors $[V_g f_p(-aK_0,bl),\ldots,
V_g f_p(aK_0,bl)]^T$, $\abs{l}\leq L_0$, are all jointly sparse with
$\lceil 2\mu^{-1} \rceil N$ nonzero coefficients. Let $\Sp$ denote
the index set of nonzero coefficients. For $\abs{\ell}\leq L_0$, let
$\bz^{\Sp}[l]$ be vectors with coefficients $z^{\Sp}_{k,l}$ defined
by
\begin{equation*}
z^{\Sp}_{k,l} = \left \{ \begin{array}{ll} z_{k,l} & k\in\Sp \\
0 & k\notin \Sp\,. \end{array} \right.
\end{equation*}
Then $\bz^{\Sp}[l]$ is the best $\lceil 2\mu^{-1} \rceil N-$term
approximation of $\bz[l]$, for each $\ell$. Note that
$\abs{z_{k,l}-z^{\Sp}_{k,l}}\leq \abs{V_g (f-f_p)(ak,bl)}$ for all
$k$ and $\ell$, so that
\begin{align*}
\norm{\bz&-\bz^{\Sp}}_{2,1} = \sum_{k=-K_0}^{K_0} \left (
\sum_{l=-L_0}^{L_0} \abs{z_{k,l}-z^{\Sp}_{k,l}}^2 \right )^{1/2} \nonumber \\
&\leq \sum_{k=-K_0}^{K_0} \left ( \sum_{l=-L_0}^{L_0}
\abs{V_g(f-f_p)(ak,bl)}^2 \right )^{1/2}  \nonumber \\
&\leq  \sum_{k=-K_0}^{K_0}  \norm{V_g(f-f_p)(ak,\cdot)}_{\lt} \leq
\sqrt{K} \norm{V_g(f-f_p)}_{\lt} \nonumber \\
&\leq \sqrt{K} C_{a,b} \norm{g}_{\SO} \norm{f-f_p}_2 \leq \delta_W \sqrt{K}
C_{a,b} \norm{g}_{\SO} \norm{f}_2\,,
\end{align*}
completing the proof.
\end{proof}


\bibliographystyle{IEEEtran}
\bibliography{Gabor_MWC}

\end{document}